\documentstyle[aps,prl,epsfig]{revtex}

\def\a{\hat{a}}
\def\A{{\cal{A}}}

\def\braket#1#2{ \langle{#1}|{#2}\rangle }
\def\opket#1#2 {  {#1} |{#2}\rangle }
\def\braopket#1#2#3{ \langle{#1}| {#2} |{#3}\rangle }

\def\c{\xi}

\def\d{{\rm d}}

\def\det{{\rm det}\,}

\def\ds{\displaystyle}

\def\eps{\epsilon}

\def\ha{{1 \over 2}}
\def\H{{\cal H}}
\def\hh{\hat{h}}

\def\I{{\cal I}}

\def\K2{{\cal K}}
\def\ket#1{ |{#1}\rangle }

\def\Khat{{K}}
\def\kt{\tilde{k}}

\def\lin{{\rm lin}}

\def\mat#1#2#3#4{  \left( \matrix{ {#1} & {#2} \cr
                                   \noalign{\vskip3pt}
                                   {#3} & {#4} \cr    } \right) }

\def\mod{{\rm mod}}

\def\phat{\hat{p}}

\def\ph#1{\phantom{#1}}

\def\qhat{\hat{q}}

\def\Re{{\rm Re}}

\def\T{\hat{\cal T}}

\def\Tim{{\cal F}}
\def\Tp{{T}}
\def\Tr{{\rm Tr}}
\def\Trl{F}

\def\Uh{\hat{U}}

\def\Wt{{\cal W}}
\def\x{{\bf x}}
\def\spinor#1#2{\left( \begin{array}{c} {#1}\\
			                {#2} \end{array} \right)}

\begin{document}

\title{Scarring and the statistics of tunnelling}
\author{Stephen C. Creagh${}^a$, Soo-Young Lee${}^a$ 
and Niall D. Whelan${}^b$ }
\maketitle
\begin{center}
{\it
$^{a}$School of Mathematical Sciences, University of Nottingham, 
Nottingham NG7 2RD, UK.
\newline
$^{b}$Department of Physics and Astronomy, McMaster University,
Hamilton, Ontario, Canada L8S~4M1.
}
\end{center}

\begin{abstract}
We show that the statistics of tunnelling can be dramatically affected
by scarring and derive distributions quantifying this effect. Strong
deviations from the prediction of random matrix theory can be
explained quantitatively by modifying the Gaussian distribution which
describes wavefunction statistics. The modified distribution depends
on classical parameters which are determined completely by
linearised dynamics around a periodic orbit. This distribution
generalises the scarring theory of Kaplan [Phys. Rev. Lett. {\bf 80},
2582 (1998)] to describe the statistics of the components of the
wavefunction in a complete basis, rather than overlaps with single
Gaussian wavepackets. In particular it is shown that correlations in
the components of the wavefunction are present, which can
strongly influence tunnelling-rate statistics. The resulting
distribution for tunnelling rates is tested successfully on a 
two-dimensional double-well potential.

\end{abstract}


\section{Introduction} 
\label{introduction}

\noindent
Tunnelling rates from systems with complex internal behaviour are
naturally described statistically. Such analyses have been used in
particular to understand nuclear resonances \cite{nuke}, chemical
reaction rates \cite{polik,miller,chemists,HOCLmeas,broad,HOCL,N2O}
and quantum dots \cite{Qdots1,Qdots2}. While the manner in which
internal states couple to the continuum varies in these analyses, the
internal quantum mechanics is usually modelled using one of the
standard ensembles of random matrix theory (RMT).  We show here that
in systems where the complexity of the internal dynamics derives from
low-dimensional chaos, sufficiently strong deviation from the standard
RMT statistics is possible that it dominates the statistics of
tunnelling. These deviations were pointed out in \cite{ourstats} and
are explained in detail here using the theory of scarring developed by
Heller, Kaplan and coworkers \cite{hellerscars,Kaplanicity}.

We consider tunnelling to and from regions of phase space associated
with chaotic behaviour. In particular, our theory works for the
calculation of level splittings in double well potentials and of
resonance widths of quasibound states in metastable wells.  With the
development of experimental techniques capable of measuring
state-specific reaction rates, such tunnelling from highly-excited
chaotic states has become
directly relevant to the analysis of chemical reactions, for example.
Specific examples of reactions which have been treated using
RMT-based statistical methods include the dissociation rates of
${\rm D_2CO}$ \cite{polik,miller} and ${\rm N_2O}$ \cite{N2O}.  
In \cite{ourstats}, the distributions arising from such an analysis were
shown to be determined in a simple way by the stability and action of
a complex tunnelling orbit which crosses the potential barrier with
minimum imaginary action. The resulting statistical distributions for
the tunnelling rate agree well with numerically computed ensembles
{\it except} when the real extension of the tunnelling orbit into the
potential well is periodic; in that case,
strong deviations from the RMT prediction 
are observed and it was proposed in \cite{ourstats} that these are due to
the effect of scarring on wavefunction statistics as outlined in
\cite{Kaplanicity}. Additional evidence in support of this has
subsequently been provided in \cite{BKH}.

In this paper we derive a distribution describing scar-influenced
tunnelling-rate statistics in two-dimensional potentials. It is 
determined completely by three dynamical parameters which are calculated 
from the monodromy matrices of the tunnelling orbit and of its real 
periodic extension. These three parameters can be understood as the 
stability of the scarred orbit, the stability of the complex tunnelling 
orbit and an angle relating stable and unstable manifolds. The root of 
this calculation is a conjecture governing the statistics of the 
components of chaotic wavefunctions in the eigenbasis of a tunnelling 
operator which was defined in \cite{ourAP}. The low-lying states of 
this tunnelling operator are approximated by the eigenstates of a 
harmonic oscillator.  Following the linear scarring theory of Kaplan 
\cite{Kaplanicity} we show that the components of chaotic states in 
this basis have variances which deviate from those of RMT and which 
must furthermore be correlated. We conjecture that the eigenfunction 
distribution is a nonisotropic Gaussian. This is the simplest 
distribution consistent with the observed correlations and may also 
be obtained from maximum-entropy arguments. This conjecture is tested 
against quantum-map models and found to describe statistics of their
eigenstates very accurately. The conjecture is then used to derive 
modified distributions for tunnelling rates in the presence of scarring 
and these are found to describe well the statistics of energy-level 
splittings in chaotic double well potentials.

We see the benefits of this work as being twofold. First, there is a solution 
to the primary problem of incorporating the effects of scarring into the 
statistics of tunnelling. This enables the detection and interpretation of 
system-specific dynamical detail in measurements of multidimensional tunnelling
and furthermore provides a very direct manifestation of the phenomenon 
of scarring in quantities with real physical relevance. Second, we believe 
that from the point of view of scarring alone, even without reference to 
tunnelling,
the correlated joint probability distribution proposed here for the statistics
of eigenstates holds considerable promise as a theoretical tool.
 In particular, it suggests methods of quantifying the effects of 
scarring in a way that does not depend  on a choice of test states. This second
aspect is not fully exploited in the present paper, but we believe that the 
essential elements needed for a full development are put in place.
We remark that the effect of correlations on the statistics of
tunnelling rates was also considered in a somewhat different context
in \cite{Qdots2}.

An outline of the paper is as follows. We begin in section~\ref{theory}
with a review of the existing theory of the statistics of tunnelling
rates.  This includes a definition of the tunnelling operator
introduced in \cite{ourAP} and a summary of the derivation from it of
nonscarred tunnelling-rate distributions as outlined previously in
\cite{ourstats}.  In section~\ref{overlapsec} we use the linear theory
of scarring to deduce that in the presence of scarring the components
of chaotic eigenstates in an eigenbasis of the tunnelling operator
must be nonRMT and conjecture a modified Gaussian distribution for
them. This conjecture is tested on quantisations of perturbed cat maps
and found to work well.  In section~\ref{pofysec} we use the
conjecture to deduce modified tunnelling-rate distributions in the
presence of scarring and these are compared successfully with
distributions of splitting in chaotic double-well potentials.

\section{Tunnelling statistics without scarring} 
\label{theory}
We begin with a brief description of existing theory of tunnelling-rate
statistics in the absence of scarring, on which our present work is 
based. We first describe a tunnelling operator defined in 
\cite{ourAP} which relates tunnelling rates to the properties
of wavefunctions in the allowed region. We then outline how this
tunnelling operator is used to derive distributions for tunnelling
rates in the absence of scarring.

\subsection{The tunnelling operator}
A full description of the semiclassical matrix element and complete
definitions of the various objects needed to calculate it are
described in \cite{ourAP}. For the purposes of statistical analysis,
an idealised model formulated in terms of quantum maps suffices, which
we summarise here.
\\
Our model starts with a finite-dimensional Hilbert space $\H$. We take
this space to be the quantum analog of a Poincar\'e section $\Sigma$ in a
potential well \cite{bogomolnytrans}, with the dimension $N$ of $\H$
being approximately proportional to the area of $\Sigma$.
A quantum mapping acts on the space $\H$ as
a unitary operator $\Uh$, whose eigensolutions we denote by
\[
\Uh\ket{n} = e^{-i\theta_n}\ket{n}
\]
and whose classical analog is a symplectic map $\Trl$ corresponding to
chaotic motion within the well. In \cite{ourAP}, the eigenstates
$\ket{n}$ are represented as certain 
%
%
%
%
cross-sections of the eigenfunctions of the usual Hamiltonian in the full 
Hilbert space.

Tunnelling rates are calculated using a {\it tunnelling operator} $\T$
which also acts on $\H$ and from which the scaled tunnelling rate
associated with a chaotic eigenstate $\ket{n}$ is calculated from
\begin{equation}\label{matel}
y_n = \frac{ \braopket{n}{\T}{n}}{\Tr\T}.
\end{equation}
In this formula, we normalise the chaotic eigenstates according to
$\braket{n}{n}=N$ so that, in particular, the overlaps with a second
basis $\ket{\kt}$ of $\H$ satisfy
$\langle|\braket{\kt}{n}|^2\rangle=1$.  As a result, $\langle
y_n\rangle=1$. In resonance problems, this scaled tunnelling rate
is  $y_n=\Gamma_n/\langle\Gamma_n\rangle$ where $\Gamma_n$ is the
resonance width associated with a particular metastable state labeled
by $n$ and in double-well problems $y_n=\Delta E_n/\langle\Delta
E_n\rangle$ where $\Delta E_n$ is the splitting of a doublet labeled by
$n$.

The tunnelling operator $\T$ is interpreted as an evolution operator
whose classical correspondent is a symplectic map $\Tim$ which is {\it
complex}.  This map $\Tim$ has a real fixed point $\zeta_0$ which
corresponds to the most probable tunnelling route. That is, using
$\zeta_0$ as an initial condition, a complex trajectory of imaginary 
period can be traced out which crosses a potential barrier with minimum 
imaginary action. Mappings of points in a neighbourhood of $\zeta_0$ in 
the Poincar\'e section $\Sigma$ correspond to complex trajectories in 
full phase space which cross the potential barrier near this central 
tunnelling route.  The matrix element in (\ref{matel}) then samples the 
state $\ket{n}$ in a small region of $\Sigma$ surrounding $\zeta_0$ and 
with area of $O(\hbar)$ (as determined by say, a Wigner function on 
$\Sigma$ \cite{ourAP}). This region is determined by the dynamics of 
$\Tim$ near $\zeta_0$ and outside of it the symbol of $\T$ decays 
exponentially. For this reason it is consistent within semiclassical 
approximation to linearise dynamics around $\zeta_0$ and let $\T$ be the 
quantisation of the resulting complex symplectic matrix $W$. Note that
in practical terms $W$ is simply the monodromy matrix of the tunnelling orbit.

We now describe the tunnelling operator for resonance problems of symmetric 
double wells where the symmetry is a reflection $(x,y)\mapsto(-x,y)$.
If we choose a representation in which $\zeta_0$ is at the centre of
coordinates $\zeta=(q,p)$ on $\Sigma$, we may write 
\[
\T = e^{-\alpha_0 \hh/\hbar},
\]
for these cases, where $\hh$ is quadratic in the corresponding operators
$\hat{\zeta}=(\qhat,\phat)$ and $\alpha_0>0$ . The corresponding
symbol $h(q,p)$ is a quadratic form in $(q,p)$, expressed as
\[
h(q,p) = \ha \zeta^T \Khat \zeta,
\]
where $\Khat$ is a positive-definite symmetric matrix.
The matrix $\Khat$ is determined from $W$ by writing
\[
W = e^{-i\alpha_0 J\Khat}.
\]
where $J$ is the unit symplectic matrix.   We normalise
$\Khat$ so that $\det\Khat=1$ and then the parameter $\alpha_0$
is fixed by a calculation of $W$.
That $W$ can be written in
this way for real $\alpha_0$ and $\Khat$ can be shown using the 
conjugate-time-reversal symmetry
\[
W^* = W^{-1},
\]
which is a special case of a similar symmetry of $\Tim$ \cite{ourAP}.  This
symmetry is a reflection of the fact that the tunnelling orbit is
periodic with an imaginary period and so complex conjugation is
equivalent to time-reversal. Note that this is the case even if the
problem does not have a proper time-reversal symmetry (such as in the
presence of magnetic fields, for example).

We limit the detailed discussion in this paper to two-dimensional 
potentials, so that the Poincar\'e section $\Sigma$ has one degree 
of freedom. In that case we may denote the eigensolutions of $\hh$ by 
\[
\hh\ket{\kt} = \left(k+\ha\right)\hbar \ket{\kt},
\]
using a tilde to distinguish them from the chaotic eigenstates
$\ket{n}$. Note then that the eigensolutions of $\T$ are then
\[
\T\ket{\kt} = e^{-(k+1/2)\alpha_0} \ket{\kt} 
= \frac{1}{\Lambda^{k+1/2}}\ket{\tilde{k}},
\]
where $\Lambda=e^{\alpha_0}$. Notice that $\Lambda$ is the larger of 
the two eigenvalues $e^{\pm\alpha_0}$ of $W$ (which are real even though 
$W$ is complex). Finally, we note that in double well problems
for which the symmetry is an inversion $(x,y)\mapsto(-x,-y)$, the eigenvalues
of $W$ as defined in \cite{ourAP} are negative and the eigensolutions
of the tunnelling operator are of a similar form except that the eigenvalues 
are $\Lambda^{-k}\left|\Lambda\right|^{-1/2}$.

\subsection{Statistics using standard RMT}
A standard statistical model of eigenstates of chaotic systems is that
their components in a generic basis are Gaussian-distributed. This
assumption has been combined with various models of coupling to the 
continuum \cite{miller,Qdots2,ourstats} to produce statistical 
distributions for the scaled tunnelling rates in the absence of 
scarring. The discussion in \cite{ourstats} forms the basis for
our treatment of scarred distributions and we therefore summarise it 
here. We assume for ease of presentation that the system is time-reversal
symmetric so the assumptions of the GOE are adopted.  Results for GUE
systems are derived similarly and are summarised at the end.

We expand the eigenstate $\ket{n}$ in an eigenbasis of $\T$,
\[
\ket{n} = \sum_k\; x_k \,\ket{\tilde{k}},
\]
suppressing the dependence of the coefficients $x_k$ on the
chaotic-state index $n$. The normalisation of $\ket{n}$ is such that
\[
\left\langle \left|x_k\right|^2 \right\rangle = 1
\]
and in systems with time-reversal symmetry, the $x_k$'s are real. The
standard RMT model is that for large $N$, the $x_k$'s are uncorrelated
and distributed with a joint probability distribution
\begin{equation}\label{RMTdis}
P(\x) =  \prod_{k=0}^{N-1} \left[\frac{1}{\sqrt{2\pi}} 
\;e^{-x_k^2/2} \right]
= \frac{1}{(2\pi)^{N/2}} \,e^{-\x^T \x/2}, 
\end{equation}
where $\x=(x_0,x_1,x_2,\cdots)$. Given that (\ref{matel}) 
expresses each $y$ as a quadratic form
\[
y = \x^T \Tp \x
\]
in $\x$, where $\Tp$ is an $N\times N$ matrix representing $\T/\Tr\T$
in a $\ket{\tilde{k}}$-basis, we can write the distribution for $y$ 
in the form
\begin{equation}\label{givepofy}
p(y) = \int 
\;\delta\!\left(y- \x^T \Tp \x \right)P(\x)\,\d\x.
\end{equation}
Fourier-transforming with respect to $y$ gives the characteristic
function,
\begin{eqnarray}\label{charint}
\tilde{p}(q) &=& \int_{-\infty}^\infty
e^{iqy}\,p(y)\,\d y\nonumber\\[3pt]
&=&  \frac{1}{(2\pi)^{N/2}}  \int
\exp\left[ -\ha\x^T(I-2iq\Tp)\x\right] \,\d\x
\nonumber\\[3pt]
&=& \frac{1}{\sqrt{\det\left(I-2iq\Tp\right)}}.
\end{eqnarray}
One finds similarly that
\begin{equation}
\tilde{p}(q) = \frac{1}{\det\left(I-iq\Tp\right)}
\end{equation}
in the absence of time-reversal symmetry.

These distributions have been shown in \cite{ourstats} to describe
successfully the tunnelling rate statistics of chaotic double wells
when the tunnelling route has a nonperiodic real extension. When the
tunnelling route connects to a periodic orbit, however, strong
deviations are found \cite{ourstats,BKH} which we explain in the next 
sections using the idea of scarring.

\section{Wavefunction statistics and scarring}
\label{overlapsec}
It has been pointed out by Kaplan and coworkers that the Gaussian
distribution in (\ref{RMTdis}) does not describe wavefunction statistics 
if a  basis state $\ket{\kt}$ is localised in phase space 
near a periodic orbit of the real classical dynamics (corresponding to the 
Poincar\'e map $\Trl$). In tunnelling-rate statistics, this can lead to 
strong deviations from the distribution described the previous section 
if $\zeta_0$, in addition to being a fixed point of the complex map $\Tim$, 
is a fixed point or short periodic orbit of $\Trl$. In this section we 
review elements of that argument and use them to calculate correlations 
among the components of $\x$. From this analysis there
emerges a natural conjecture for generalising (\ref{RMTdis}). 
The distributions we calculate on the basis of this conjecture are necessary 
to understand tunnelling but also, independently of the tunnelling problem, 
hold promise as a basis for understanding the statistics of scarring in a 
rather general way.

\subsection{A Gaussian hypothesis}
For a given ensemble of states with overlaps $\x=(x_0,x_1,x_2,\cdots)$, 
let us define a matrix of correlations $C$ whose elements are
\begin{equation}\label{defC}
C_{lk} = \langle x_l x_k^* \rangle.
\end{equation}
The average is over the eigenstates $\ket{n}$ and will be made more
precise below.  We will in particular consider ensembles formed by the
eigenstates $\ket{n}$ of $\Uh$ whose eigenangles lie in a subset of
the unit circle. By varying the quantum dimension $N$, for example, we
can consider ensembles corresponding to fixed classical maps $\Trl$
and $\Tim$. If the joint probability distribution is of the form
(\ref{RMTdis}) or its GUE equivalent, we expect
\begin{equation}\label{CisI}
C=I,
\end{equation}
which says simply that $\langle|x_k|^2\rangle=1$ and that $x_k$ and 
$x_l$ are uncorrelated unless $k=l$. In order to avoid confusion
with a correlation function defined below, we will henceforth refer 
to $C$ as the {\it covariance matrix}. We now show that deviations
of $C$ from the identity can be calculated on the basis of linearised 
dynamics around the point $\zeta_0$. The construction begins with a 
calculation of quantum recurrence for basis states localised near
$\zeta_0$ and is finished by a Fourier transformation which relates 
these to the averages $\langle x_l x_k^* \rangle$.

Let $\{\ket{\kt}|\;k=0,1,\cdots\}$ be an eigenbasis of $\T$ as defined 
in the previous section and suppose that $\zeta_0$ is a fixed point of 
the classical map $\Trl$ corresponding to $\Uh$. We now argue that for 
some ensembles (\ref{CisI}) cannot hold. Following \cite{Kaplanicity} 
we define the correlation function
\begin{eqnarray}\label{Aexp}
A_{lk}(t) &=& \braopket{\tilde{l}}{\Uh^t}{\tilde{k}} \nonumber\\[6pt]
	  &=& \frac{1}{N}\sum_{n=1}^N  x_l x_k^*\,e^{-i\theta_nt}
\end{eqnarray}
and denote its Fourier transform by
\begin{eqnarray}\label{Sexp}
S_{lk}(\theta) &=& \sum_{t=-\infty}^\infty e^{i\theta t} A_{lk}(t) 
							\nonumber\\[6pt]
&=& \frac{2\pi}{N}\sum_{n=1}^N x_l x_k^*\; \delta(\theta-\theta_n).
\end{eqnarray}
The correlation function is approximated semiclassically by a sum of the form
\begin{equation}\label{scA}
A_{lk}(t) \approx \sum_p \A_p e^{2\pi NiS_p}
\end{equation}
over classical orbits $p$, typically complex, which begin in and
return to a small region around $\zeta_0$ after $t$ iterations of
$\Trl$, assuming $k$ and $l$ are $O(1)$. Each orbit contributes a term
with an amplitude $\A_p$ which is independent of $N$ and an exponential
$\exp[2\pi NiS_p]$ with a rapidly-varying complex phase which decays
exponentially with the distance of the contributing orbit from
$\zeta_0$. The primary such orbit corresponds to the fixed point
$\zeta_0$ itself. We are free to choose the phase of $\Uh$ so that the
action of $\zeta_0$ vanishes (this amounts to a choice of the origin
of the $\theta_n$ values). With that choice it is manifest that its
contribution to $A_{lk}(t)$ is $N$-independent.

Consider averaging $A_{lk}(t)$ over a sequence of quantum systems with
varying $N$ but identical classical limits. We find then that
\[
\left\langle A_{lk}(t) \right\rangle 
\approx \sum_p \A_p \left\langle e^{2\pi NiS_p}\right\rangle 
\]
and, with the exception of the fixed point $\zeta_0$ for which $S_p=0$,
the contributions average to zero. The contribution from $\zeta_0$ can 
be obtained from linearised dynamics and we find that
\[
\left\langle A_{lk}(t) \right\rangle  \approx A^\lin_{lk}(t) 
\equiv\braopket{\tilde{l}}{\Uh_\lin^t}{\kt}
\]
where $\Uh_\lin$ is the quantisation of a linearisation of the real
map $\Trl$ at $\zeta_0$. The linearised correlation function
$A^\lin_{lk}(t)$ is calculated in the next section. It is
$N$-independent and decays exponentially to zero with time when 
$k$ and $l$ are $O(1)$.

Performing the same averaging procedure on $S_{lk}(\theta)$ gives
\[
\left\langle S_{lk}(\theta) \right\rangle \approx S^\lin_{lk}(\theta)
\equiv\sum_{t=-\infty}^\infty e^{-i\theta t} A^\lin_{lk}(t).
\]
This is a real, $2\pi$-periodic function of $\theta$ which is also
$N$-independent. Suppose we form an ensemble of chaotic states by
varying $N$ and selecting states for which $\theta_n$ lies within a
narrow window centred about $\theta$. The eigenfunction
expansion in (\ref{Sexp}) indicates that
\begin{equation}\label{Clin}
C_{lk}(\theta) = \left\langle x_lx_k^* \right\rangle = 
\left\langle S_{lk}(\theta) \right\rangle \approx S^\lin_{lk}(\theta).
\end{equation}
We find therefore that, if we produce an ensemble of overlaps
from states within a narrow window of phase angle, the covariance
matrix violates condition (\ref{CisI}) and the joint probability 
distribution for $\x$ necessarily deviates from the standard 
{\rm RMT} form.

The simplest distribution consistent with (\ref{Clin}) is the Gaussian 
\begin{equation}\label{gausshyp}
P(\x;\theta) = \frac{1}{\left[(2\pi/\beta)^N\det C(\theta)\right]^{\beta/2}}\;
e^{-\beta\x^T C^{-1}(\theta)\,\x/2}
\end{equation}
where $\beta=1$ in the GOE case and $\beta=2$ in the case of GUE.  
We conjecture that these distributions
govern the wavefunction statistics in the presence of scarring when
states are taken from a small window of eigenphases. A full
justification of this would require a detailed analysis of long orbits
contributing to (\ref{scA}), along the lines of the nonlinear theory
of scarring in \cite{Kaplanicity}. We will not offer such an analysis
here and assume (\ref{gausshyp}) on the grounds of simplicity alone. This 
may be formalised by noting that (\ref{gausshyp}) minimises the 
information content \cite{Balian,Mehta}
\[
{\cal I}\left[ P(\x)\right] = \int P(\x)\ln P(\x) \; \d\x,
\]
subject to the constraint that the correlations are as written in
(\ref{defC}) and that $N$ is large. Minimisation of information is not
a proof, however, and ultimately we rely on detailed numerical testing
to justify the choice of distribution.
We now show in detail how $C(\theta)$ may be calculated
and verify that the Gaussian hypothesis describes quite well the 
scarred wavefunction statistics of specific quantum-map models.

\subsection{Calculation of the covariance matrix}
Construction of the scarred distribution in (\ref{gausshyp})
begins with the calculation of the linearised correlation function
$A_{lk}^\lin(t)$, from which the covariance matrix $C(\theta)$ is 
obtained by Fourier transformation. We show in this subsection how
$A_{lk}^\lin(t)$ may be calculated from the monodromy matrices
linearising real and complex dynamics around the fixed point
$\zeta_0$ (corresponding to the maps $\Trl$ and $\Tim$ respectively). 
In the main text we present enough detail that the method
of calculation of $A_{lk}^\lin(t)$ should be clear and leave the
derivation and other detailed discussion to appendices.  Once
$A_{lk}^\lin(t)$ has been constructed as outlined below, the covariance
matrix is then easily calculated in practice using a FFT, for 
example.

It is shown in appendix~\ref{appgetA} that $A_{lk}^\lin(t)$ vanishes
if $l$ and $k$ are not both even or both odd and may otherwise be 
calculated from the polar form
\begin{equation}\label{givecorr}
A_{k+2n,k}^\lin(t) =  G_{kn}(\psi(t)) \;
e^{i(k+1/2)(\phi(t)-\mu t\pi)+in\phi(t)+ in\pi/2},
\end{equation}
where the Maslov index $\mu$ and the angles $\psi(t)$ and $\phi(t)$ 
are defined by dynamics around $\zeta_0$ as described below.
The amplitude is
\[
G_{kn}(\psi) = \sqrt{\frac{k!}{(k+2n)!}}\;\frac{(2n)!}{2^nn!}\;\;
\sin^n\psi \,\sqrt{\cos\psi} \; C_k^{n+1/2}(\cos\psi),
\]
where $C_k^\alpha(x)$ denotes a Gegenbauer polynomial. In the special 
case $n=0$ we get the autocorrelation function
\[
A_{kk}^\lin(t) = \sqrt{\cos\psi(t)} \; P_k (\cos\psi(t))\; 
e^{i(k+1/2)(\phi(t)-\mu t\pi)},
\]
where $P_k (x)$ is a Legendre polynomial.
We assume $n\geq 0$ and $t>0$ in (\ref{givecorr}) and use 
${A_{lk}^\lin}^*(-t)=A_{lk}^{\lin}(t)=A_{kl}^\lin(t)$
to calculate the correlation function when $t<0$ or $l<k$.
Note that this assumes a convention for the phases of the 
eigenstates $\ket{\kt}$ that is outlined at the end of 
appendix~\ref{appgetrace}. In particular, this convention 
leads to a covariance matrix $C(\theta)$ which is real-symmetric 
even in the case of GUE statistics; that is, while individual values 
of $x_lx_k^*$ are complex, the phase convention is such that the
averages  are real.

To define the angles $\phi(t)$ and $\psi(t)$, let $M_0$ be the real
symplectic matrix linearising the unstable real dynamics around
$\zeta_0$ and let
\begin{equation}\label{defW}
W(z) = e^{i(\ln z) J\Khat}
\end{equation}
be a generalisation of the complex symplectic matrix described in the
the last section, reducing to that case when $z=e^{-\alpha_0}$.
Euler expansion of the exponential in (\ref{defW}) leads to the identity
\[
\Tr\, W(z)M_0^t = (-1)^{\mu t}\left(m(t)z + \frac{m^*(t)}{z}\right),
\]
where
\[
m(t) = \cosh \rho t + iQ\sinh\rho t
\]
and where $(-1)^\mu e^{\pm\rho}$ are the eigenvalues of $M_0$. The
real parameter $Q$ follows the notation of \cite{Kaplanicity} and is 
given by $Q=\cot\varphi$, where $\varphi$ is the angle between the
stable and unstable manifolds of $M_0$. Note that the parameter $Q$ 
depends implicitly on $W$ because the angle $\varphi$ is calculated 
using the metric defined on phase space by $\Khat$. Finally, 
we complete the explanation of (\ref{givecorr}) by defining the angles 
$\psi(t)$ and $\phi(t)$ implicitly 
using the polar decomposition 
\[
m(t) = \sec\psi(t) \,e^{i\phi(t)}
\]
of $m(t)$. We may choose these angles to lie in the ranges
$0<\psi(t)<\pi/2$ and $-\psi(t)<\phi(t)<\psi(t)$ respectively 
(note that $\sec\psi\cos\phi = \cosh\rho t>1$) 
and, with this restriction on $\phi(t)$, the half-angle in
(\ref{givecorr}) is well-defined.

Note that, as $t\to\infty$, we find 
\[
\cos\psi(t) \sim \frac{2}{\sqrt{1+Q^2}}\,e^{-\rho t}
\]
and the correlation function in (\ref{givecorr}) therefore decays at
the exponential rate $e^{-\rho t/2}$ when $k$ is even and at the rate
$e^{-3\rho t/2}$ when $k$ is odd (since $C_k^{n+1/2}$ is then an odd
polynomial). Less obvious from (\ref{givecorr}) is the fact that the
correlation function also decays for fixed $t$ with increasing $k$ and
$l=k+2n$. This is a reflection of the fact that more excited states of
the oscillator $\hh$ are less localised at the fixed point $\zeta_0$
and therefore less affected by scarring.

When the linear correlation function is Fourier-transformed, we
therefore get a covariance matrix $C_{lk}(\theta)$ which can be
decomposed into two blocks corresponding to odd and even $l$ and $k$
and whose elements approach those of the identity matrix
($C_{lk}(\theta)\to\delta_{lk}$) as $l$ and $k$ increase. We expect
scarring therefore to affect the statistics of $x_k$ significantly for
relatively small values of $k$ and to have little effect when $k$ is 
large. 
%
%
This exponential convergence towards identity means that the matrix $C-I$ 
is trace-class and the determinant of $C$ exists in the limit
$N\rightarrow\infty$.  In the next subsection we calculate the
resulting covariance matrix for some model quantum maps and show that
the Gaussian hypothesis provides a good model for the
wavefunction statistics of scarred systems.

\subsection{Testing the Gaussian hypothesis}
We test the Gaussian hypothesis using a model system in which $\Uh$
quantises the perturbed cat map,
\begin{eqnarray*}
\Trl_\eps = \Trl_{0}\circ  M_{\eps},
\end{eqnarray*}
where
\[
\Trl_0 : \spinor{q}{p} \mapsto 
\mat{1}{1}{1}{2}\spinor{q}{p} \qquad\mod\;1
\]
is the standard cat map and 
\[
M_{\eps} :\spinor{q}{p} \mapsto 
\spinor{q-\eps\sin 2\pi p}{p} \qquad\mod\;1
\]
is a kicked Harper map. For sufficiently small values of the parameter
$\eps$ this map shares the hyperbolic structure of the unperturbed 
map but not its nongeneric degeneracies. Details of the quantisation
of these maps for Hilbert spaces of arbitrary dimension $N$ may be 
found in \cite{OdA,SCC}.

The map $\Trl_\eps$ has a symmetry of inversion about the 
origin, which is always therefore a fixed point. It
also has a less obvious time-reversal symmetry. 
One finds that
\[
P\Trl_\eps P = \Trl_\eps^{-1},
\]
where
\[
P : \spinor{q}{p} \mapsto 
\mat{-1}{0}{-1}{1}\spinor{q}{p} \qquad\mod\;1.
\]
Note that $P$ is antisymplectic and that $P^2=I$. The map therefore 
allows us to test statistics in the GOE scenario.
 (A systematic discussion of 
such time-reversing symmetries can be found in \cite{bris}).

To complete our model of a tunnelling system, we construct an analog
of the tunnelling operator. Consider the Harper-like Hamiltonian
\[
h(q,p) = 2-\cos \pi(2p-q)-\cos 2\pi q.
\]
This is invariant under the time-reversal symmetry $P$ and has 
a minimum at the origin in phase space. The function does not 
have the full periodicity of the torus in $q$ but, by placing the 
resulting discontinuity at $q=1/2$,  may be quantised 
in such a way that the ground and first few 
excited states of its quantisation $\hh$ are localised near the origin.
These are then approximated semiclassically by those of the harmonic 
oscillator $(\phat-\qhat/2)^2+\qhat^2$. The statistics of the components
$x_k$ of chaotic eigenstates of $\Uh$ in an eigenbasis of
$\hh$ therefore provide a test of the Gaussian hypothesis outlined
in the previous subsection and since $\Uh$ and $\hh$ share a time-reversal 
symmetry, the GOE hypothesis is appropriate.

\begin{figure}[h]
\vspace*{-0.5cm}\hspace*{0.0cm}\epsfig{figure=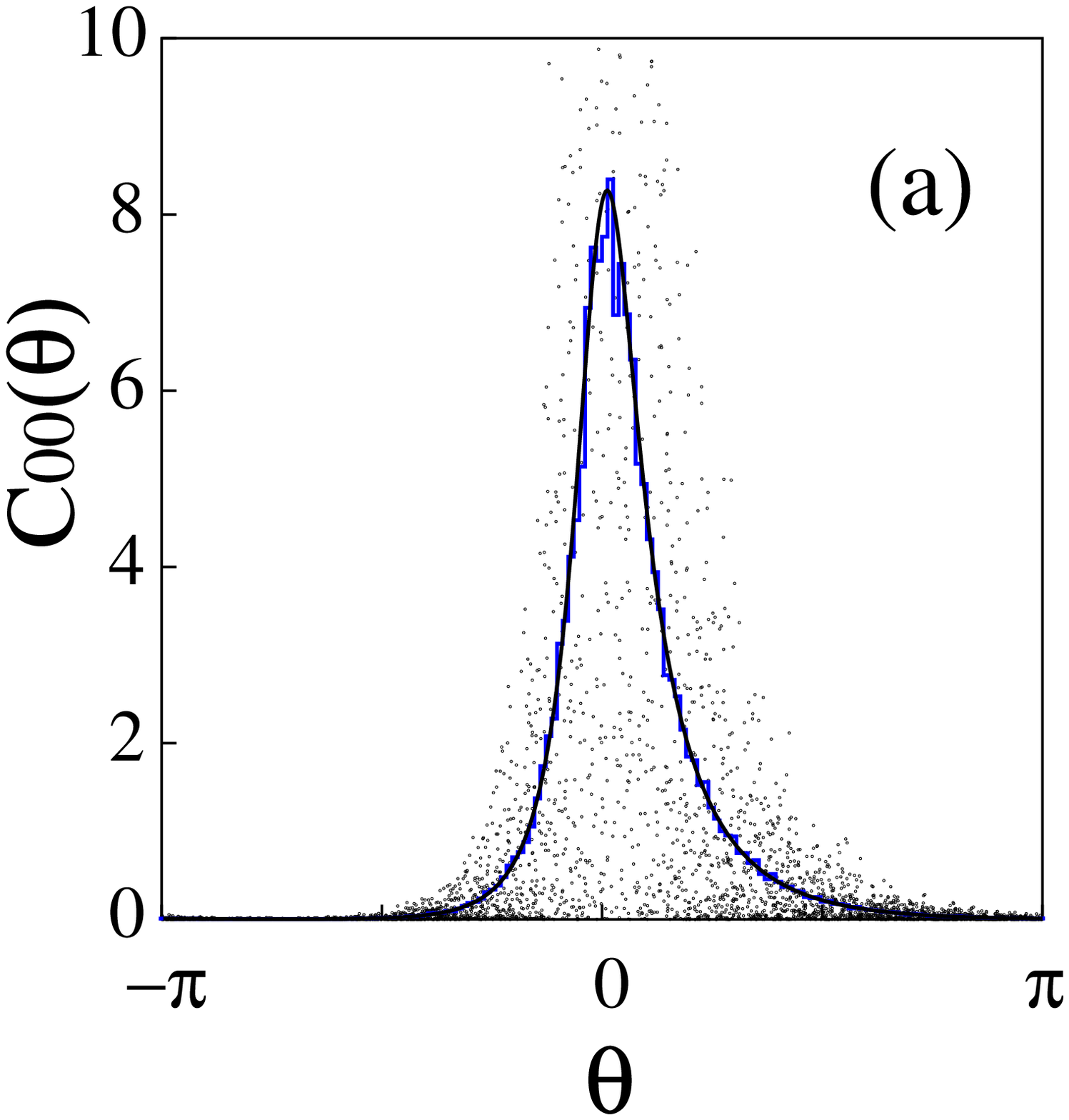,height=2.5in}

\vspace*{-2.5in} \hspace*{2.2in}\epsfig{figure=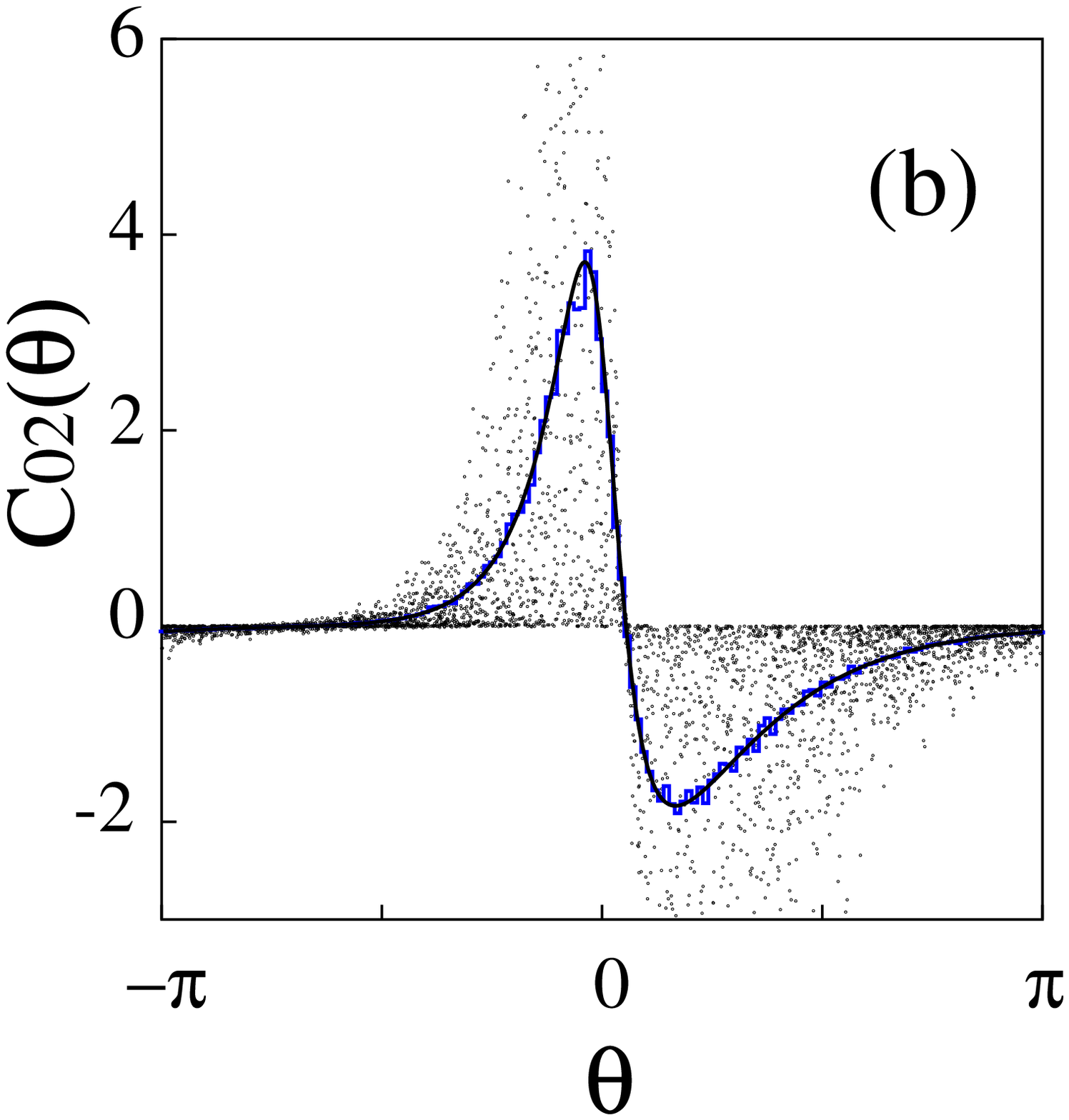,height=2.5in}

\vspace*{-2.5in} \hspace*{4.4in}\epsfig{figure=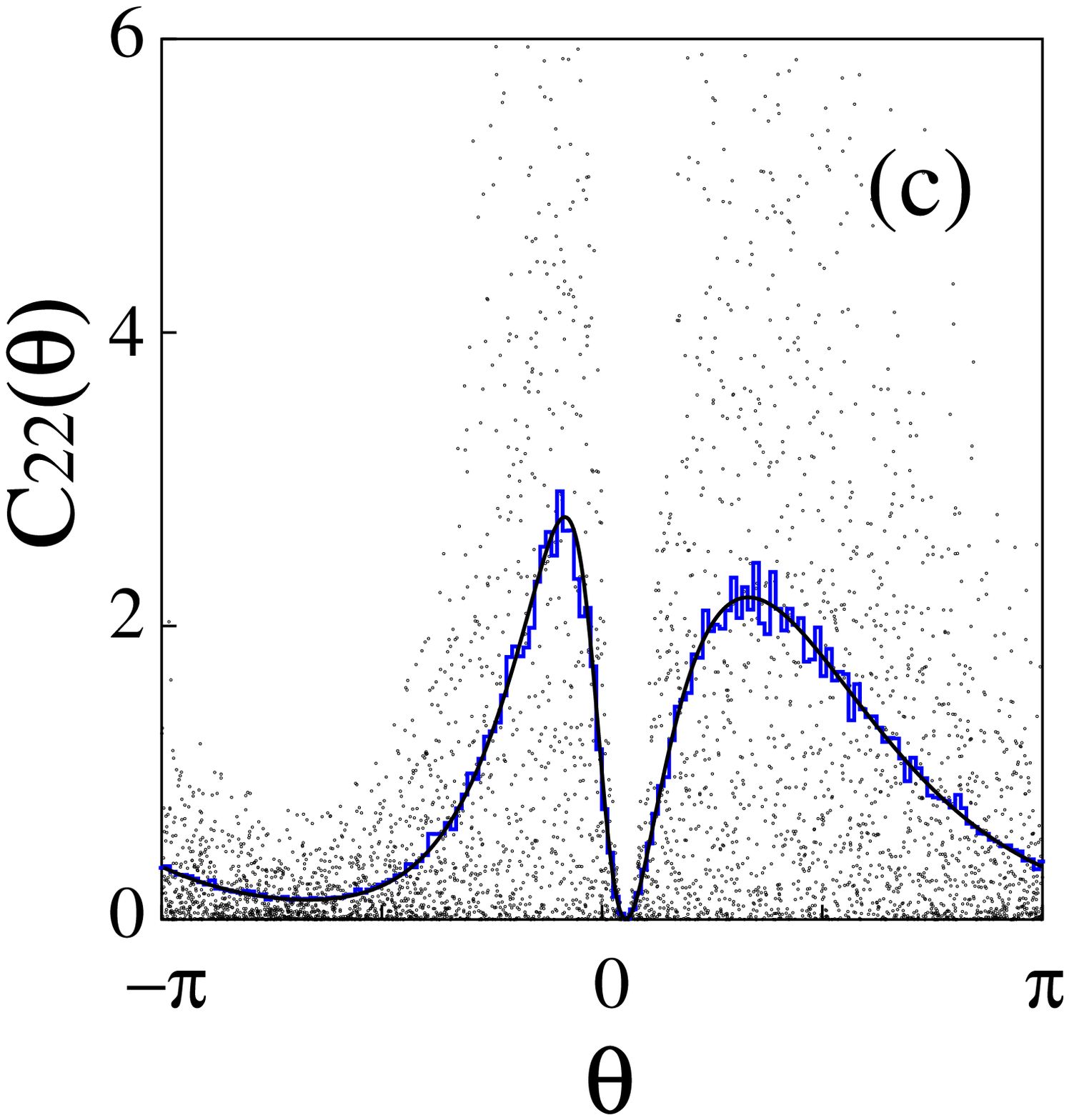,height=2.5in}

\vspace*{-0.2in}
\caption{The smooth curves are calculated using our prediction for the 
covariance matrix as a function of phase angle $\theta$. We show 
$C_{00}(\theta)$, $C_{02}(\theta)$ and $C_{22}(\theta)$ in $(a)$, $(b)$ and $(c)$ 
respectively. In each case we find good agreement with the step curve, which 
represents the local average obtained by binning components 
of the eigenstates of $\Uh$ for $\eps=0.1$ and $N$ between $100$ and 
$700$. The points represent overlaps of individual states.}
\label{localav}
\end{figure}

We first consider the prediction for the covariance matrix $C(\theta)$. The 
element $C_{00}(\theta)$, which corresponds to the overlap statistics of a 
Gaussian wavepacket (defined as the ground state of $\hh$), has been considered 
extensively by Kaplan and coworkers. For comparison with later calculation it 
is plotted as a function of $\theta$ in Fig.~\ref{localav}$(a)$ for the map 
$\Trl_\eps$ with $\eps=0.1$. 
Following \cite{Kaplanicity}, we refer to this curve as the {\it spectral envelope}. 
We expect larger than average overlaps where $C_{00}(\theta)>1$ and we will 
refer to this as the scarred region. The rest of the interval is the 
antiscarred region, where we expect smaller than average overlaps. The points 
in the figure give a representative sample of individual overlaps. There is
also a curve representing a binned average of these overlaps which is 
obtained by accumulating quantum data for even values of $N$ between $100$ 
and $700$. It is also important for our analysis to understand the
statistics of $x_lx_k$ when $l$ and $k$ lable excited states of $\hh$. We 
therefore make an analogous comparison with quantum data
for $C_{02}(\theta)$ and $C_{22}(\theta)$ in Figs.~\ref{localav}$(b)$ and 
\ref{localav}$(c)$ respectively. Note the different vertical scales in 
these cases. In all cases the average obtained by binning quantum data agrees 
well with the prediction $C_{lk}(\theta)$ obtained by Fourier-transforming 
the correlation function. This is to be expected since we have a relatively 
direct argument showing that this should be so. Less obvious is that
the distribution of individual overlaps about these averages should be 
governed by the Gaussian hypothesis, since we do not have a proof for that.
That this assumption can be verified numerically (below) is therefore 
nontrivial.

To compare the distributions about the mean $C_{lk}(\theta)$,
lets us denote
\[
\c_{lk} = \Re \,x_l x_k^* = \x^\dagger B_{lk}\x,
\]
where $B_{lk}$ is a symmetric matrix whose $ij^{\rm th}$ entry is
$(\delta_{il}\delta_{jk}+\delta_{ik}\delta_{jl})/2$. (For the moment we 
allow either GUE or GOE statistics even though all the numerical calculations 
we present are for the GOE case).
Similarly to the calculation of $p(y)$ in 
section~\ref{theory}, we find that the local distribution
$p(\c_{lk};\theta)$ is predicted by the Gaussian hypothesis to have 
the Fourier transform
\begin{eqnarray}\label{localmom}
\tilde{p}_{lk}(q;\theta) &=& \int_{-\infty}^\infty
e^{iq\c_{lk}}\,p(\c_{lk};\theta)\,\d \c_{lk}\nonumber\\[6pt]
&=& \frac{1}{\left[(2\pi/\beta)^N\det C(\theta)\right]^{\beta/2}}\;
\int
\exp\left[ -\ha\x^\dagger\!\left(C^{-1}(\theta)-\frac{2iq}{\beta}
					B_{lk}\right)\!\x\,\right] \d\x
\nonumber\\[6pt]
&=& \left[\det\ph{'\!\!}\left(I-\frac{2iq}{\beta} B_{lk}C(\theta)\right)
					\right]^{-\beta/2}.
\end{eqnarray}
The determinant simplifies on expansion to
\[
\det\left(I-\frac{2iq}{\beta} B_{lk}C(\theta)\right) 
	= 1-\frac{2iq}{\beta}C_{lk}
		+\frac{q^2}{\beta^2}\left(C_{ll}C_{kk}-C_{lk}^2\right)
\]
and Fourier transformation of (\ref{localmom}) then gives
\[
p(\c_{lk};\theta) = e^{\beta\c_{lk}C_{lk}/D_{lk}} f(\c_{lk};D_{lk},B_{lk}).
\]
Here, $D_{lk}=C_{ll}C_{kk}-C_{lk}^2$, $B_{lk} =\sqrt{C_{ll}C_{kk}}$ and 
$f(\c;D,B)$ is an even function of $\c$, defined by
\[
f(\c;D,B)     =
\cases {  
{\ds\frac{1}{\pi D^{1/2}} \;K_0\left(B\c /D\right)},
					& for GOE and   \cr
                                                   \noalign{\vskip9pt} 
{\ds \frac{1}{B}
\exp\left[-2B|\c|/D\right]
 },
					& for GUE, \cr
}
\]
(in which $K_0(z)$ is a modified Bessel function). 

\begin{figure}[h]
\vspace*{-0.5cm}\hspace*{1.0cm}\epsfig{figure=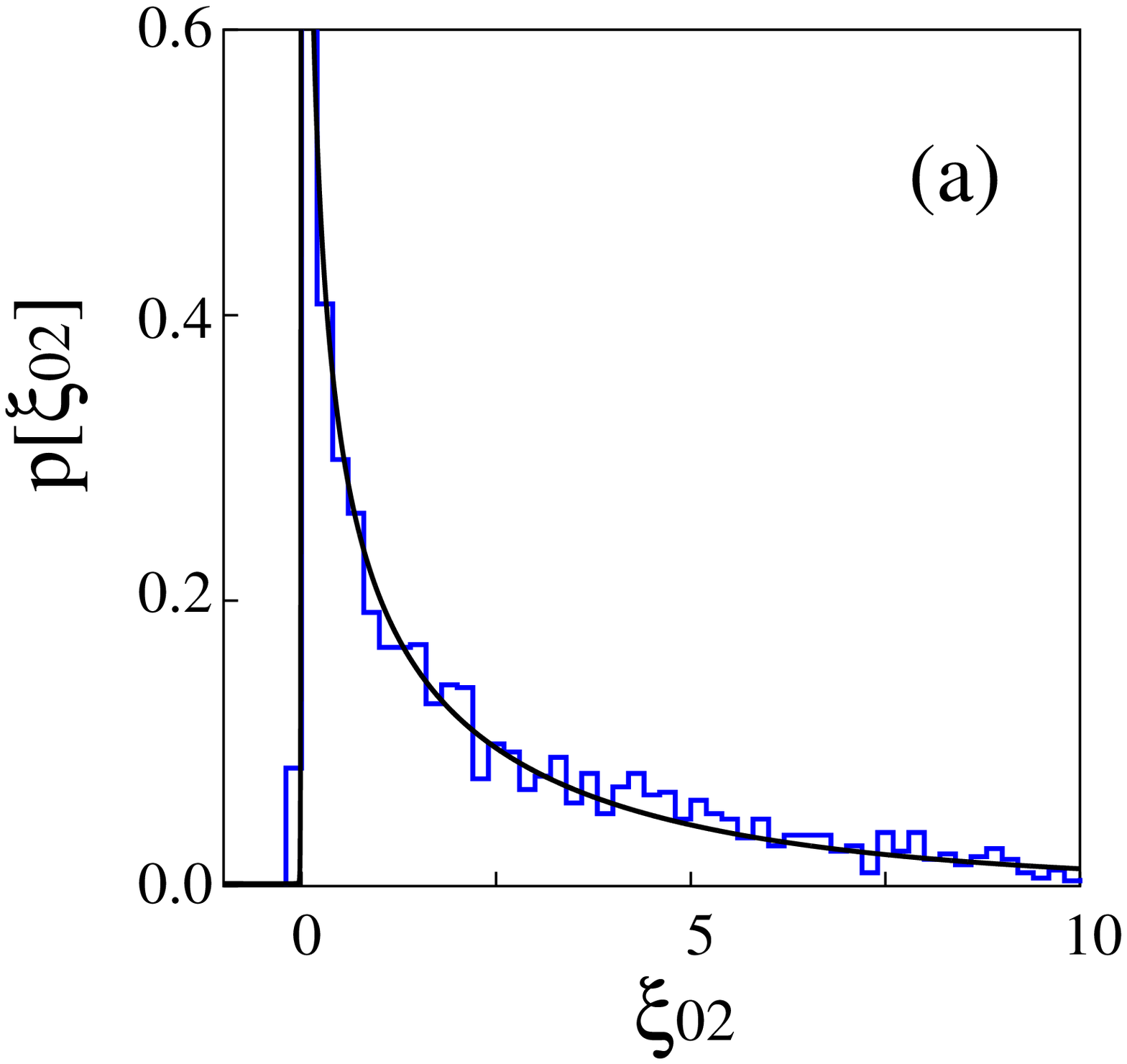,height=3.0in}

\vspace*{-3.0in} \hspace*{3.4in}\epsfig{figure=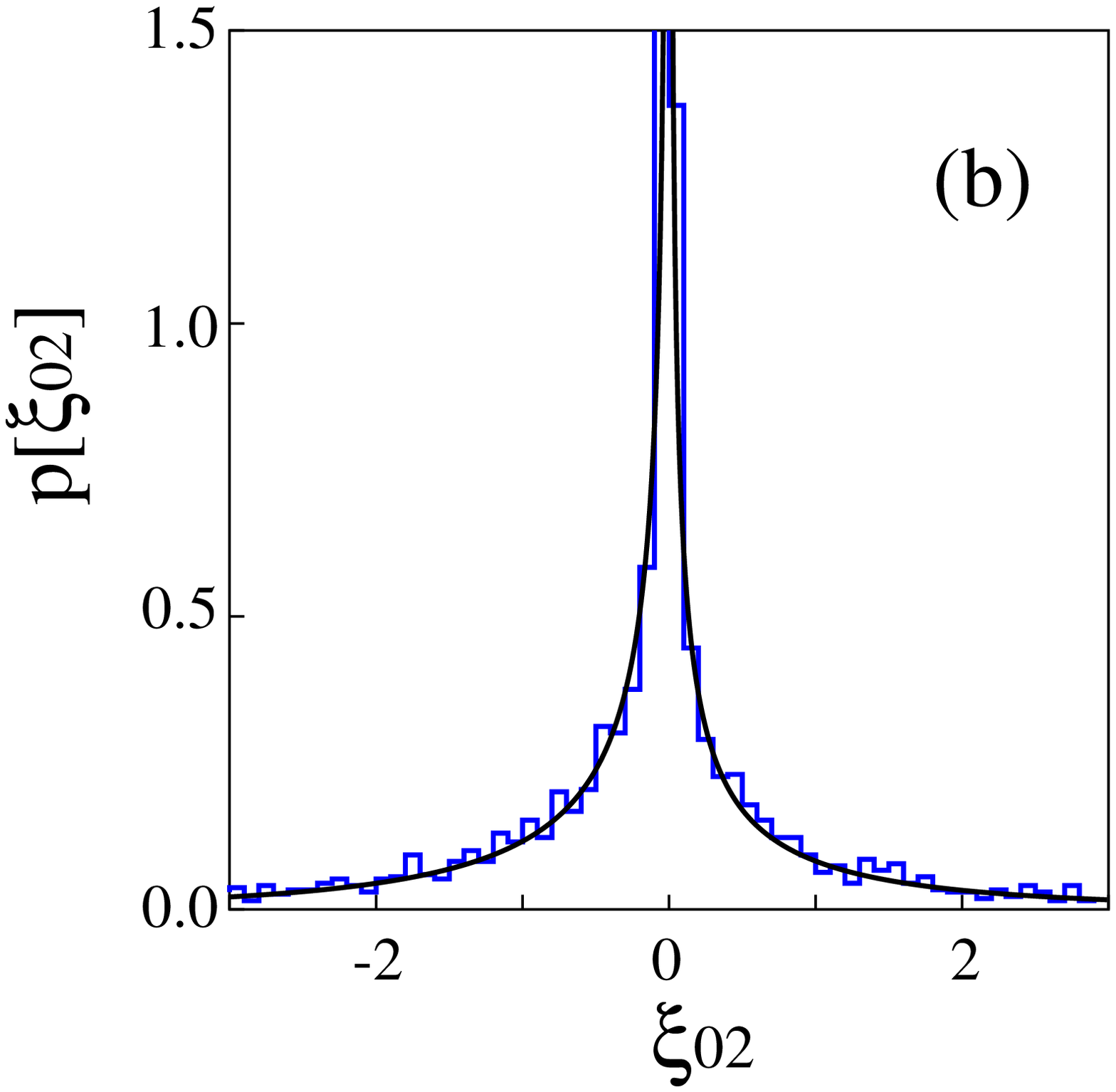,height=3.0in}

\vspace*{-1.7cm}\hspace*{1.0cm}\epsfig{figure=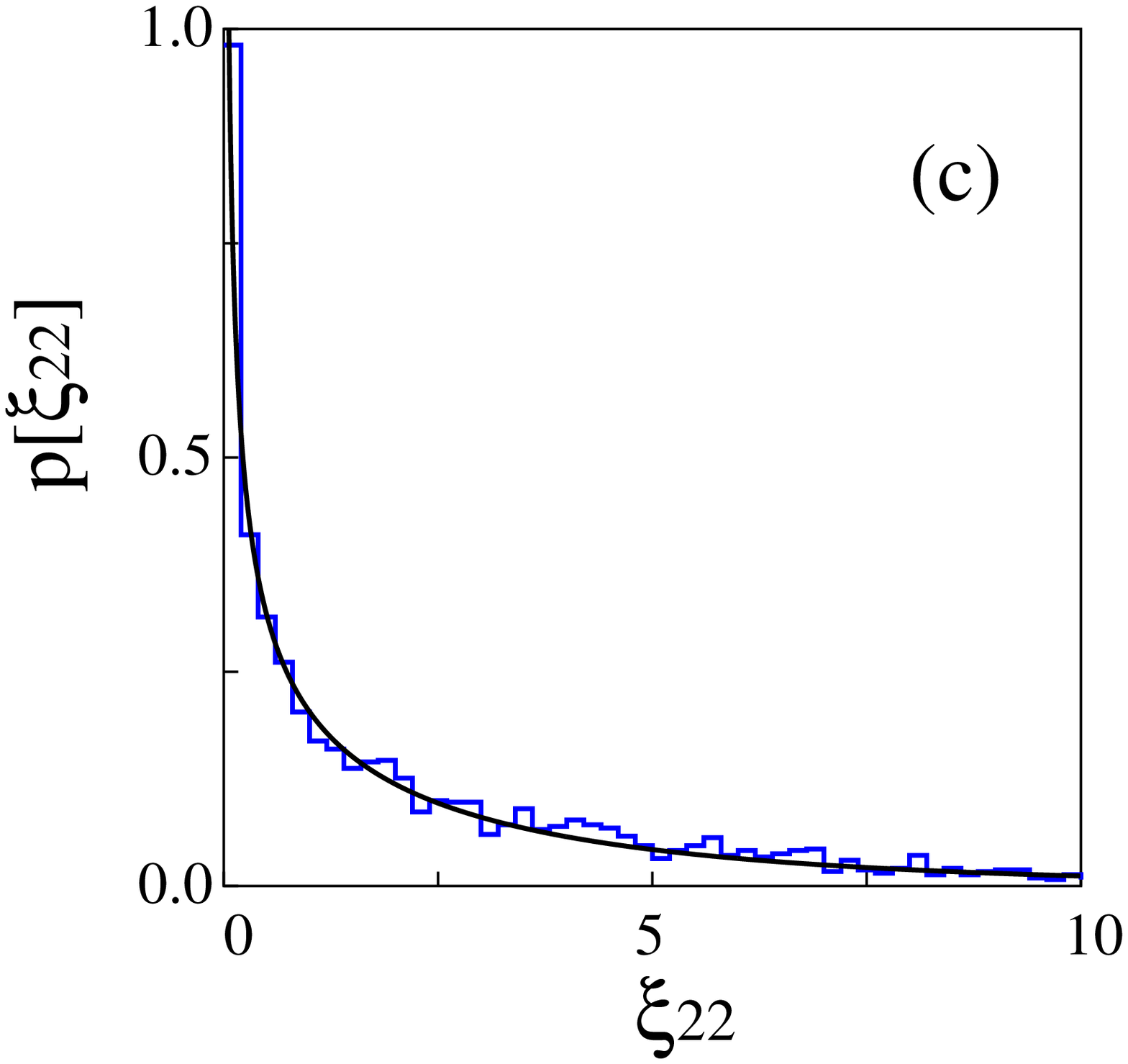,height=3.0in}

\vspace*{-3.0in} \hspace*{3.4in}\epsfig{figure=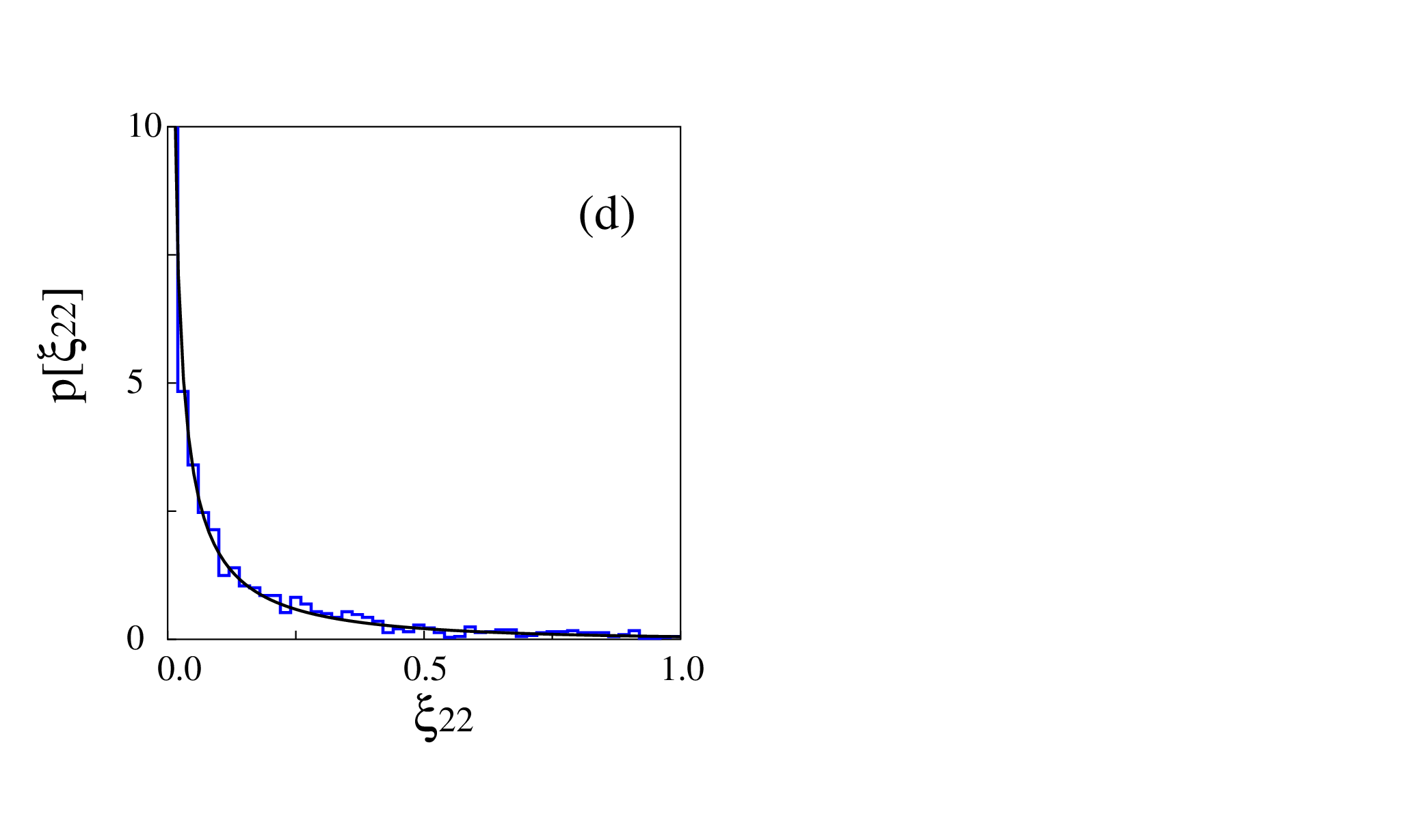,height=3.0in}

\vspace*{-0.5cm}
\caption{In (a) and (b) the statistical distribution of 
$\c_{02}=x_0x_2$ over the intervals  $\I=(-0.14\pi,-0.06\pi)$
and $\I=(0.02\pi,0.10\pi)$ respectively are compared with the
prediction of the Gaussian hypothesis. The two intervals chosen 
correspond respectively to regions of phase angle which are scarred 
and (relatively) antiscarred. Figures (c) and (d) show the statistics of 
$\c_{22}=x_2^2$ in the same scarred and antiscarred intervals. Note the 
different scales in the scarred and antiscarred cases.}
\label{subdis}
\end{figure}

When $l=k$ we find in the GOE case that 
\begin{equation}\label{ptscaled}
p(\c_{kk};\theta) = \frac{\c_{kk}^{-1/2}}{\sqrt{2\pi C_{kk}(\theta)}}\;
e^{-\c_{kk}/2C_{kk}(\theta)},
\end{equation}
which is just the Porter-Thomas distribution scaled so that
$\langle \c_{kk}\rangle=C_{kk}(\theta)$, consistent with 
\cite{Kaplanicity}.

In order to test these predictions with reasonable statistics, we create
ensembles by selecting states with phase angles restricted to subsets
$\I$ of the unit circle and compare their statistics with the
accumulated distribution
\begin{equation}\label{pI}
p_\I(\c_{lk}) = \frac{1}{\mu(\I)}\int_\I p(\c_{lk};\theta)\, \d \theta,
\end{equation}
where $\mu(\I)=\int_\I\d \theta$. These ensembles provide averages that
are local in $\theta$ while allowing sufficient data to make meaningful
numerical tests.
The resulting statistical distributions are compared in
Fig.~\ref{subdis} with the statistics of $\c_{02}=x_0 x_2$ and
$\c_{22}=x_2^2$. In each case the statistics in two subintervals are
shown: one covering a region where there is scarring (larger than
average overlaps) and one in a region where there is antiscarring
(smaller than average overlaps). In all cases the Gaussian hypothesis
works  well in describing the whole distribution. Note that similar
agreement has been found for models with GUE statistics (which are 
not shown here in the interests of brevity).
\begin{figure}
\vspace*{-0.5cm}\hspace*{0.8in}\epsfig{figure=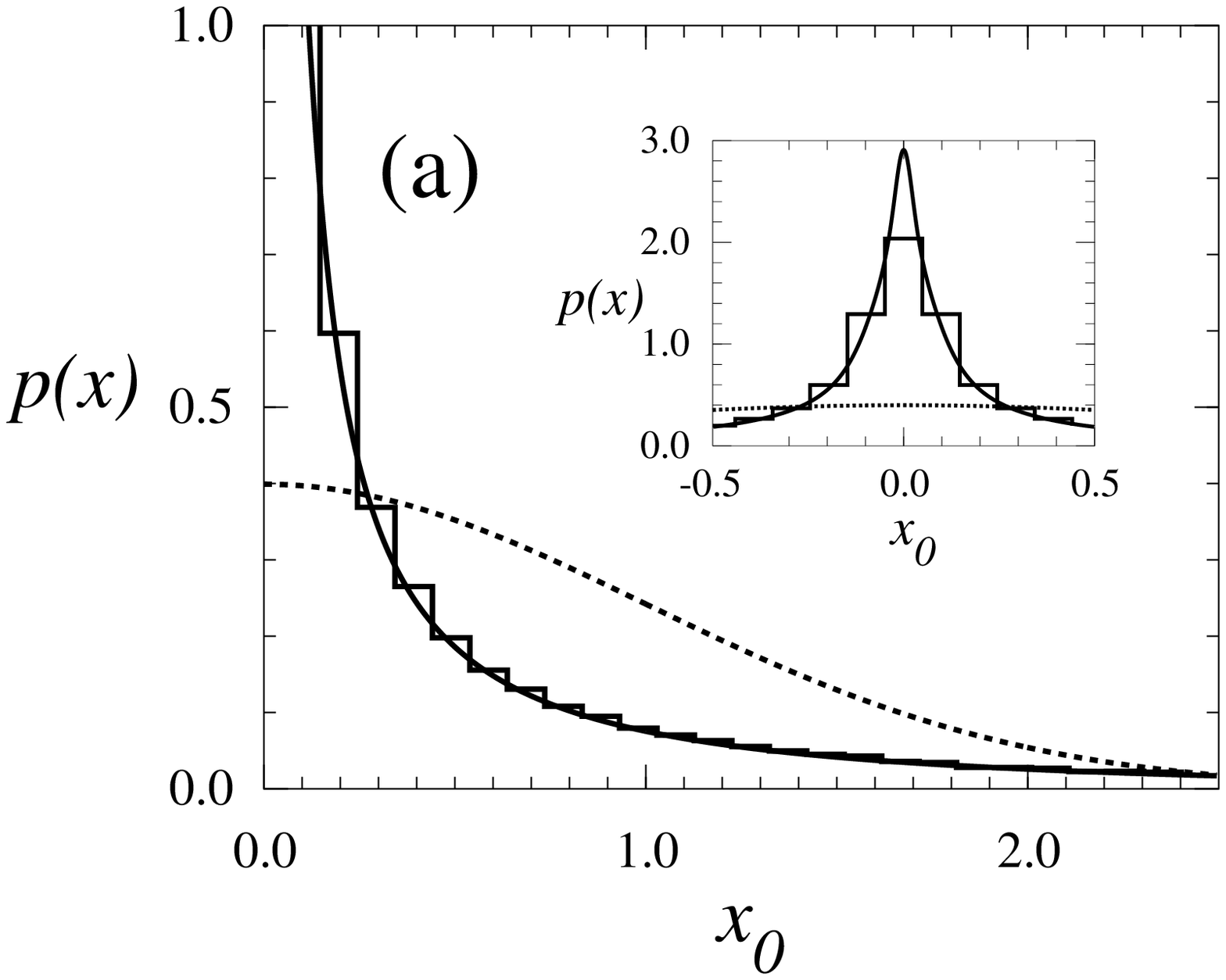,height=4.0in}

\vspace*{-4.0in} \hspace*{3.3in}\epsfig{figure=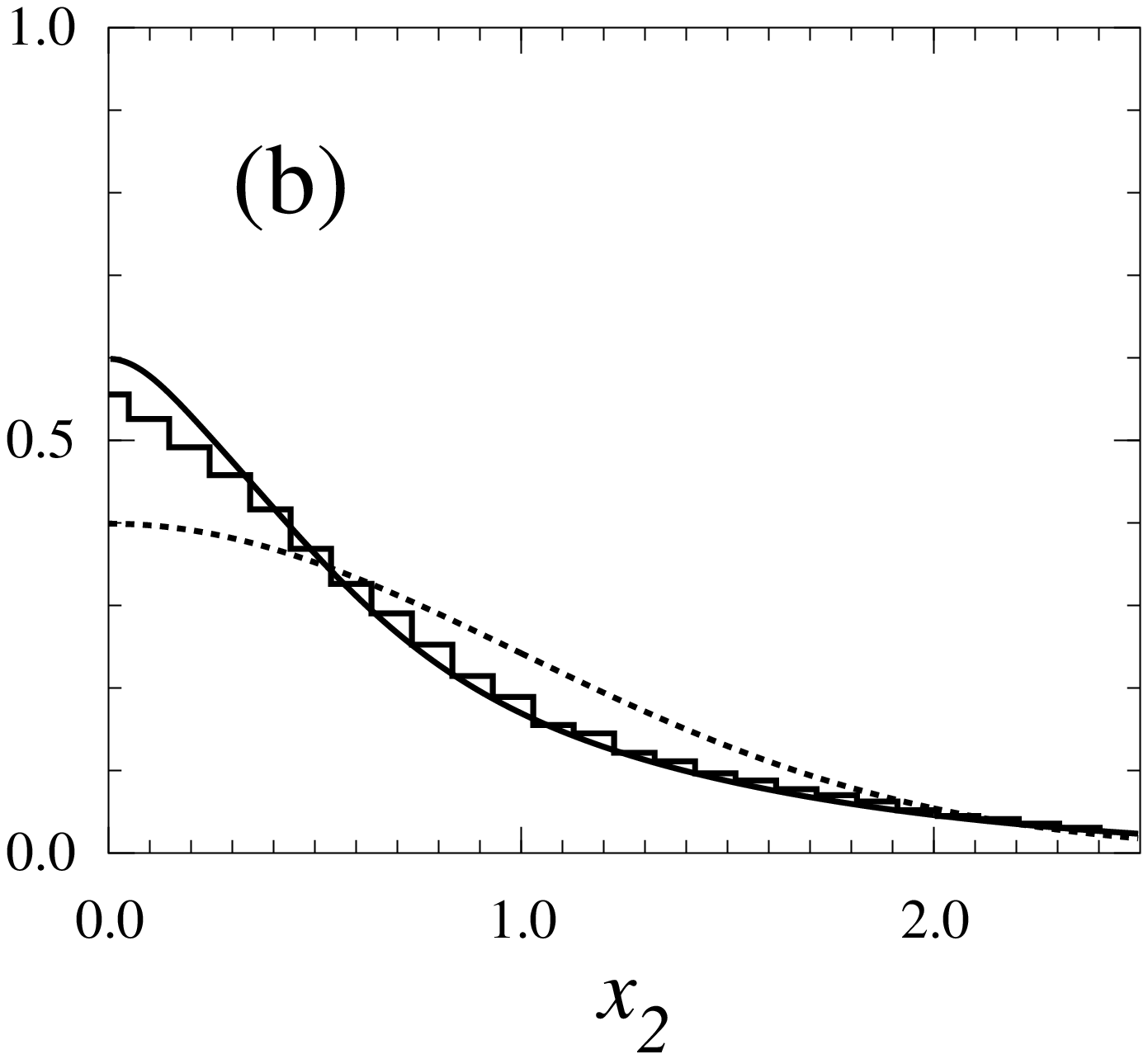,height=4.0in}

\vspace*{-5cm}\hspace*{0.8in}\epsfig{figure=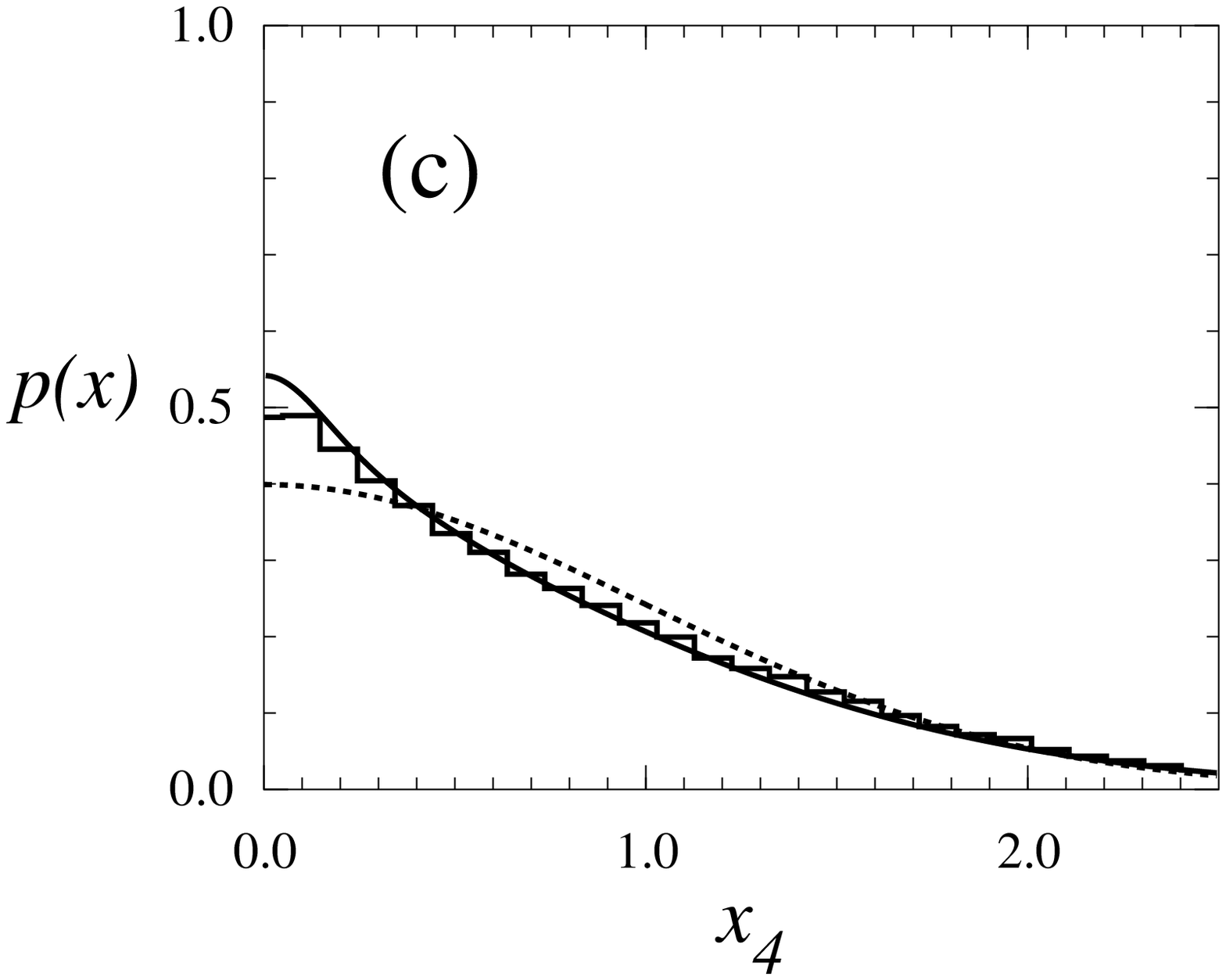,height=4.0in}

\vspace*{-4.0in} \hspace*{3.3in}\epsfig{figure=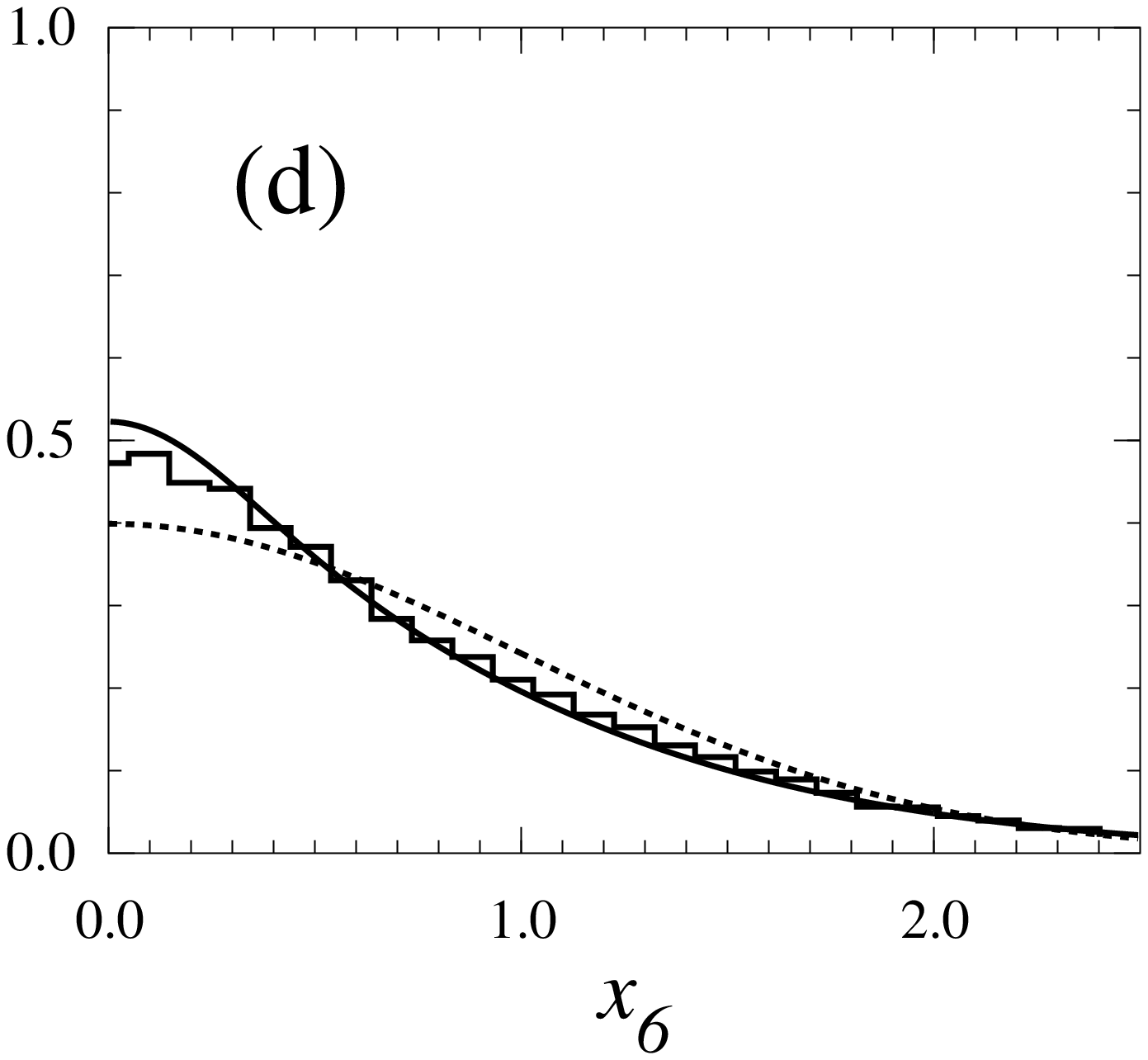,height=4.0in}

\vspace*{-4cm}

\caption{The distributions of $x_k$, accumulated over the whole
unit circle of phase angle, are shown in (a)-(d) for $k=0,2,4,6$
respectively. The insert in (a) shows the tip of the distribution.
The Gaussian hypothesis (solid curve) is in good 
agreement with the quantum data (histogram) over the whole 
distribution. In each figure the dashed curve is the normal, 
unscarred, RMT prediction. The scarred distributions approach 
this RMT case as $k$ increases (although not necessarily 
monotonically in general).}
\label{pxks}
\end{figure}
We also show, in Fig.~\ref{pxks}, distributions for the individual 
overlaps $x_k$. The local distributions $p(x_k;\theta)$ are Gaussians scaled 
to that the variances are $C_{kk}(\theta)$, analogously to (\ref{pxks}). We 
then define accumulated distributions by integrating over an interval 
$\I$ with respect to $\theta$, as in (\ref{pI}). In practice, the most 
natural ensemble is formed by averaging over the whole unit circle, 
corresponding to taking $\I=[0,2\pi)$. A sequence of such distributions 
is shown in Fig.~\ref{pxks} corresponding to $k=0,2,4,6$. One sees
that the distribution approaches the normal GOE prediction as $k$ 
increases but that for smaller values of $k$ the deviation from GOE 
is quite strong. Note that the Gaussian hypothesis describes quite 
accurately the statistics for moderate  values of
$x_k$ as well as in the tails which correspond to the strong scarring
limit. There is some underestimation in the antiscarring limit
where $x_k$ is small, but the overall agreement is  good
over the whole range of values of $x_k$. Similar or better agreement has 
been found in all the models we have examined.

All the numerical evidence we have collected indicates strongly that the 
Gaussian hypothesis accurately describes the complete statistics of chaotic 
eigenstates in a neighbourhood of a periodic orbit. While motivation 
for extending the linear theory of scarring to a complete basis has initially
come from a desire to treat tunnelling, we note that the results in this 
section are interesting from the point of view of scarring alone. The 
covariance matrix 
$C(\theta)$ provides a means of characterising the eigenstate statistics 
in the neighbourhood of a periodic orbit in a way that is independent of 
the choice of test state. We can accomodate the characterisation of 
statistics using a basis other than $\{\ket{\tilde{k}}\}$ simply by altering 
the covariance matrix using a similarity transformation. Ideally, any basis
we use should in the interests of compactness cover the immediate 
neighbourhood of the periodic orbit (as measured by Wigner functions, for 
example) with as few members as possible, but we are otherwise then 
free to manipulate those basis elements at will. In addition to being
essentially basis-independent in this way, such a characterisation
is complete, in the sense if we choose a subset of the basis which 
covers a semiclassical neighbourhood of the periodic orbit, the 
statistics of any other measure of the eigenstate near the
periodic orbit could in principle then be determined from the 
corresponding sub-block of the covariance matrix.

Note finally that the deviations from RMT of the statistics of low-lying 
components of the eigenstate persist, and in fact remain fixed, in the 
semiclassical limit. This  can be understood intuitively to derive from 
the fact that as $\hbar\to 0$ (or $N\to\infty$) the support of a harmonic 
oscillator basis state shrinks at the same rate as the semiclassical 
footprint of the periodic orbit. In $\Sigma$, the area occupied by each 
decays as $O(\hbar)$ in the semiclassical limit. This semiclassical shrinking 
of the test states we use allows the scarring effect to remain fixed
while other measures of scarring (such as the value of a wavefunction 
at a fixed point in space) lead to a vanishing effect in the semiclassical 
limit.

\section{Tunnelling statistics with scarring}\label{pofysec}
The scarred wavefunction statistics described in the previous section
are now used to derive distributions for the anomalous tunnelling
rate statistics of \cite{ourstats,BKH}. The first step is to
outline how the discussion for maps in the preceding section may be 
adapted to the statistics of states in potential-well problems
and then we can describe how these ideas are used to construct 
distributions for tunnelling rates.

While the formalism of tunnelling is  presented in \cite{ourstats,ourAP} in 
terms of quantum maps and their eigenstates, quantisation of a potential-well 
problem leads more naturally to wavefunctions in a full Hilbert space and
at first sight it is not obvious how to relate the two descriptions.
Faced with a numerically-produced
wavefunction, for example, there is no exact, purely quantum-mechanical 
way of defining a corresponding state in a finite-dimensional Hilbert 
space; all existing descriptions are semiclassical in nature. 
Our task is made easier by the fact that tunnelling depends only on 
properties of the state in a small region of phase space (or of a Poincar\'e 
section in the reduced description) surrounding a tunnelling orbit. 
Within that region, the dynamical properties of the tunnelling orbit
have been used in \cite{ourAP} to give an explicit prescription
for the definition of a Wigner function on a Poincar\'e section.
We will not repeat the details of that prescription here but state 
simply that a recipe exists for turning an eigenfunction
$\psi_n(x,y)$ of a Hamiltonian with potential $V(x,y)$ into a Wigner 
function $\Wt_n(y,p_y)$ on a Poincar\'e section $\Sigma$ defined by 
$x=x_0$ and on which $(y,p_y)$ are canonical coordinates.
This Wigner function is invariant near the tunnelling orbit under 
iteration of a semiclassical quantisation of the Poincar\'e map
\cite{bogomolnytrans}. With appropriate normalisation
this construction allows us to compute overlaps, using
\begin{equation}\label{Wigover}
\left|\braket{n}{\tilde{k}}\right|^2 = \int_\Sigma \d y \d p_y 
	\Wt_n(y,p_y) \tilde{\Wt}_k (y,p_y),
\end{equation}
between a reduced chaotic state $\ket{n}$ and 
an eigenstate $\ket{\tilde{k}}$ of a harmonic oscillator Hamiltonian 
$h(y,p_y)$ defined on the Poincar\'e section $\Sigma$.

We now check that these overlaps are described statistically
by the distributions derived in the preceding section. Until now the 
discussion has been given entirely in terms of simple quantum maps
for which the chaotic spectrum is described in terms of eigenangles 
on the unit circle. In physical problems the spectrum is more 
naturally described in terms of energy levels on the real line and we 
must connect the two pictures before proceeding. A detailed connection 
can be made using the transfer-operator approach of \cite{bogomolnytrans}, 
but for present purposes it is sufficient to identify the part of the 
spectrum corresponding to a narrow segment of the spectral envelope (so 
that we may partition states into ensembles with simple Gaussian
overlap statistics). A continuous-time analog of the phase angle
$\theta$ is provided by the action $S(E)$ of the periodic orbit
responsible for scarring \cite{Kaplanicity}. We therefore partition the 
quantum states into ensembles for which  $\theta=S(E)/\hbar$ lies in narrow 
intervals, mod~$2\pi$. We assert that the statistics of the
overlaps in (\ref{Wigover}) for each such ensemble should be a 
nonisotropic Gaussian such as described in (\ref{gausshyp}).

\begin{figure}
\vspace*{1cm}\hspace*{.3cm}\epsfig{figure=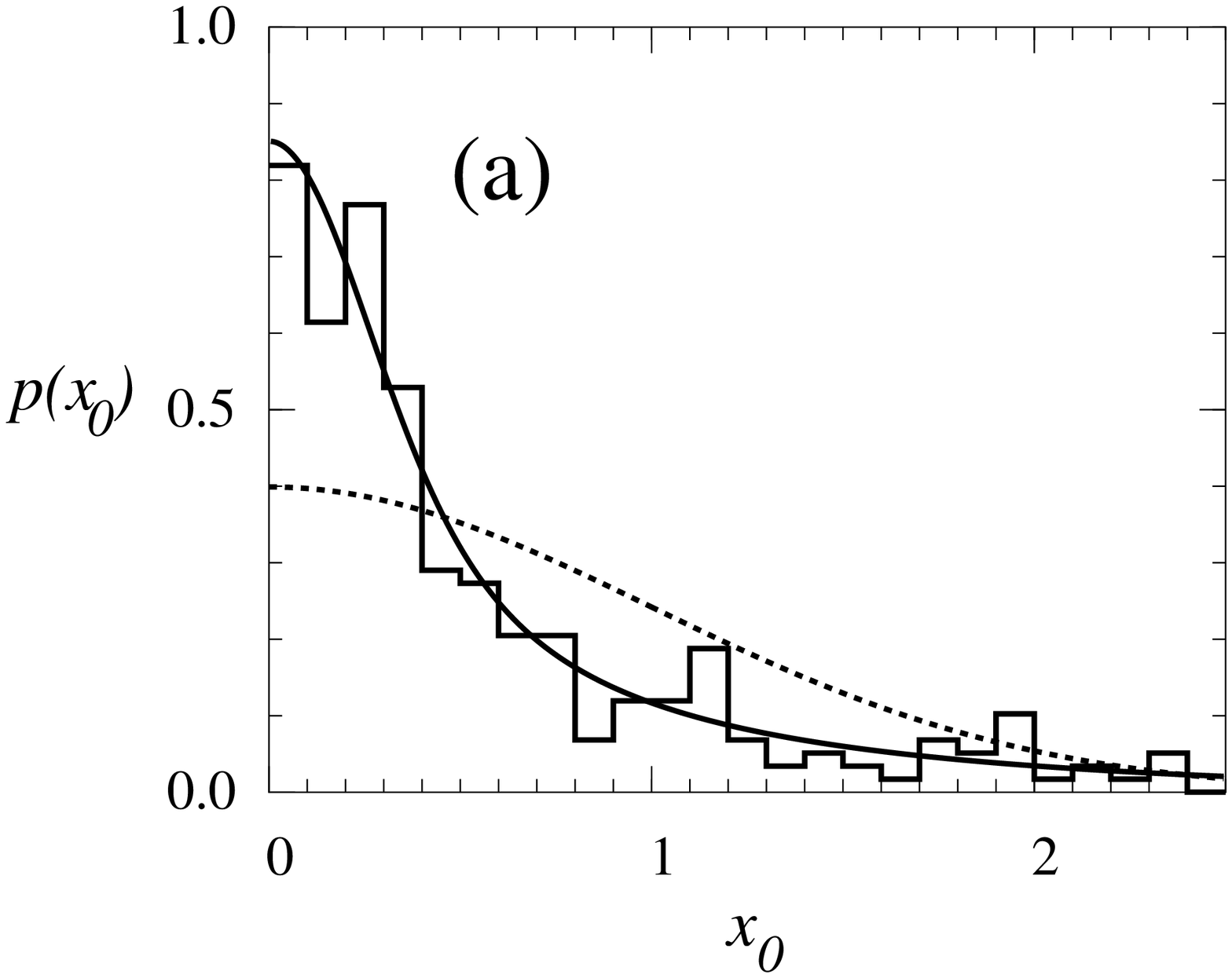,height=3.0in}

\vspace*{-3.0in} \hspace*{3.5in}\epsfig{figure=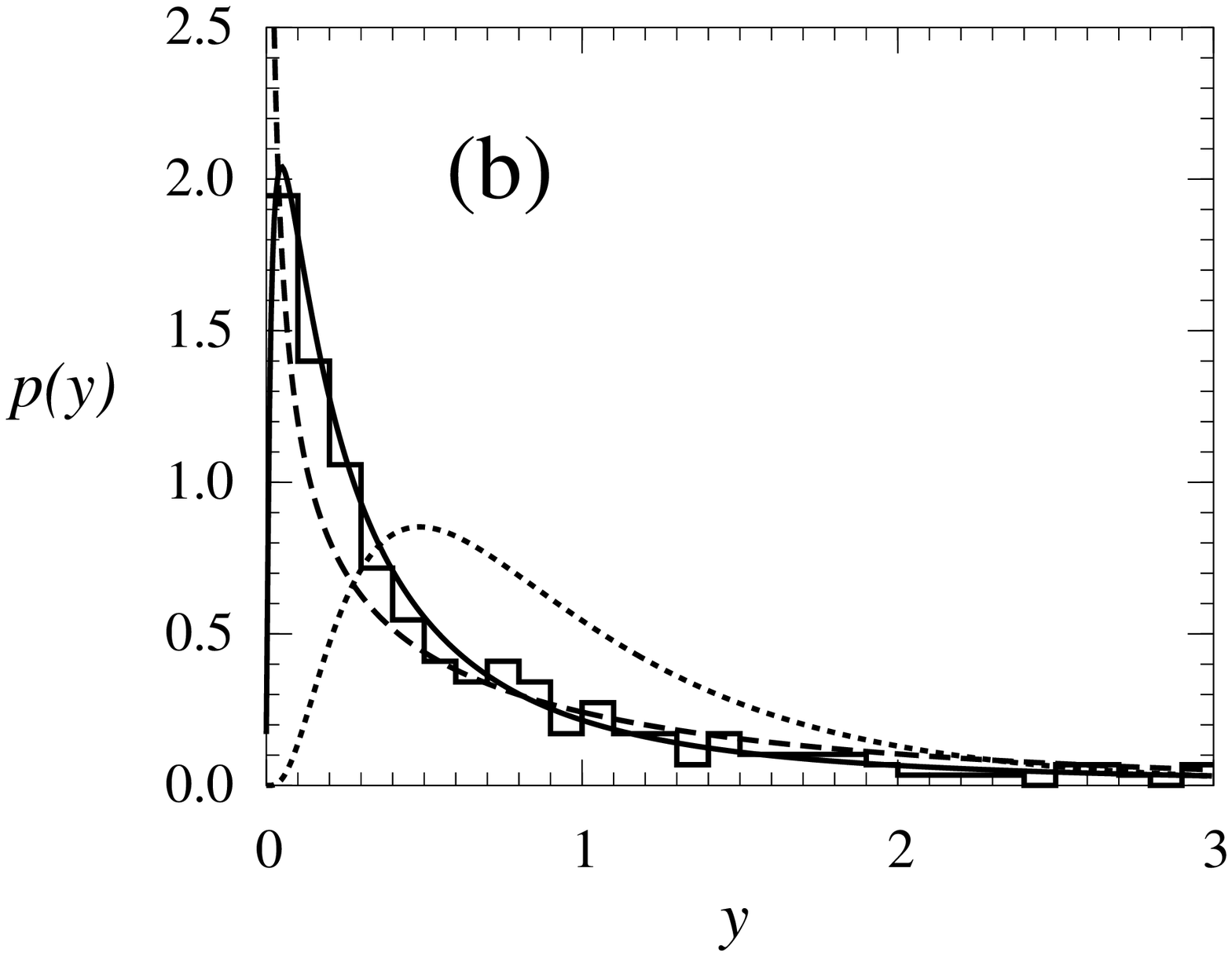,height=3.0in}

\vspace*{-.5in} \hspace*{0.3cm}\epsfig{figure=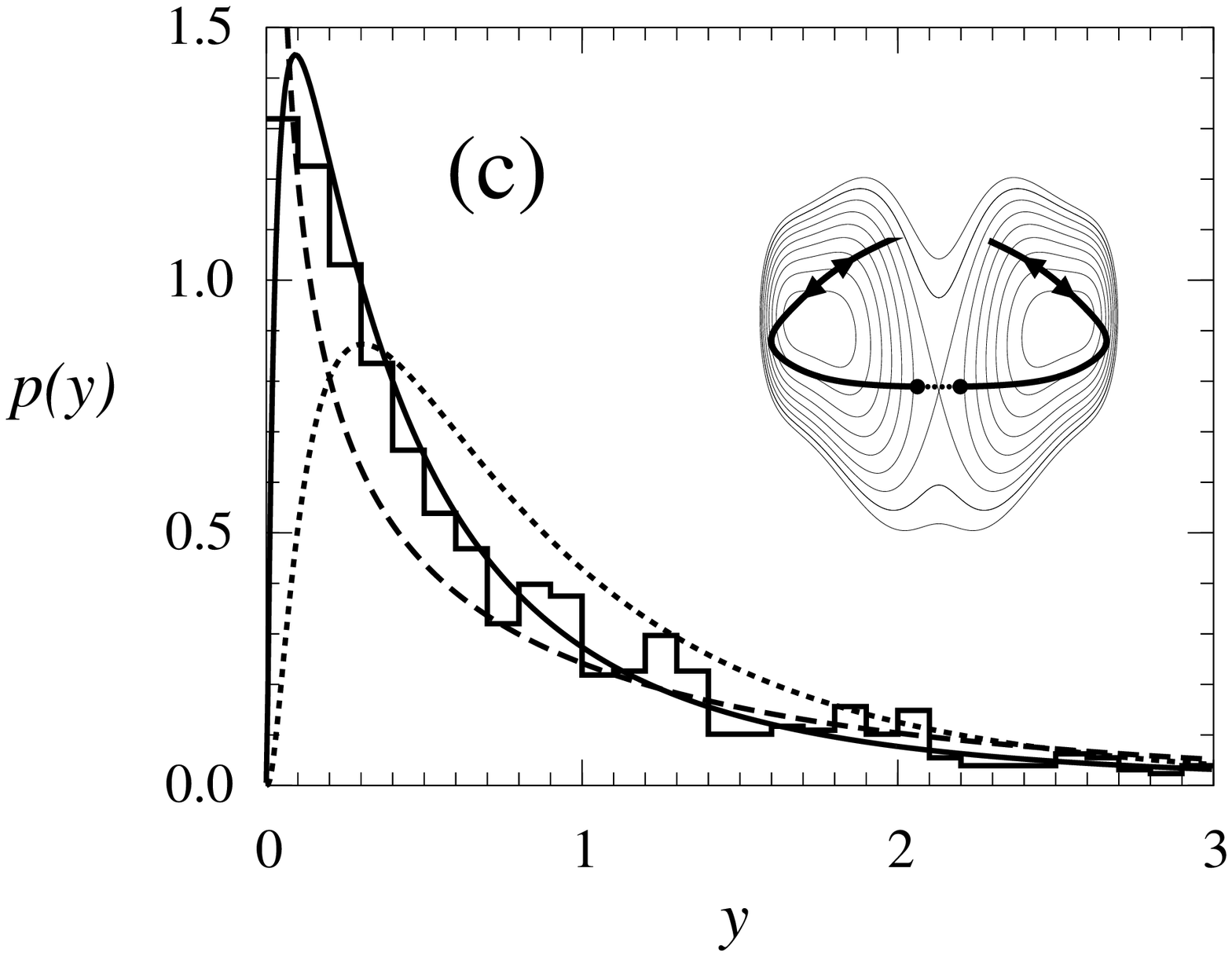,height=3.0in}

\vspace*{-3.0in} \hspace*{3.5in}\epsfig{figure=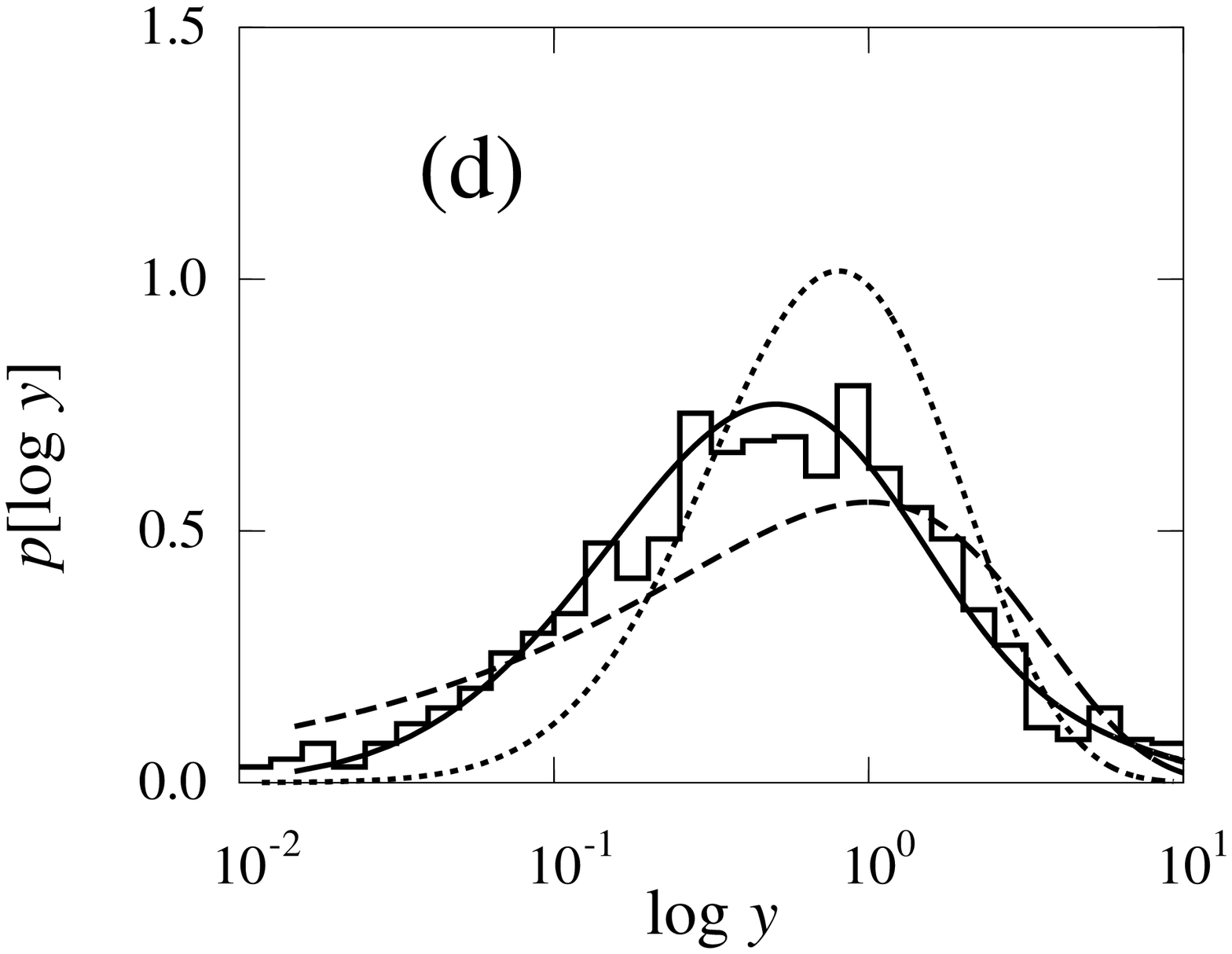,height=3.0in}

\vspace*{-2.5cm}
\caption{In $(a)$ the theoretical distribution (solid curve) 
for overlaps with a harmonic-oscillator ground state is compared 
with quantum data (histogram) for the ($y$-symmetric) potential 
with $(\mu,\nu,\sigma) = (0.1,0,0)$. The energy is fixed at the value 
$E=0.9$ (where the potential saddle has energy $E=1$) and $\hbar$ is 
quantised in the quantum calculation. The histogram represents
392 states with $48<1/\hbar<75.77$, consisting of 12 periods of the 
spectral envelop function. In $(b)$ we show the 
corresponding distribution of rescaled tunnelling rates. In addition to
the standard RMT prediction without scarring (short-dashed curve) 
we also show the Porter-Thomas distribution as a long-dashed curve. 
In $(c)$ and  $(d)$ we present tunnelling rate statistics, using linear 
and log scales respectively, for the potential with 
$(\mu,\nu,\sigma)=(0.25,0.40,0.254)$. In this case states 1480 states
with $40<1/\hbar<106.47$, representing 54 periods of the spectral 
envelop function, are used. This potential is not symmetric 
in $y$ but the real extension of the tunnelling orbit is still periodic 
(inset). As in $(a)$ and $(b)$, the short-dashed curve represents a 
RMT calculation without scarring, the solid curve the scarring distribution
and the  long-dashed curve shows the Porter-Thomas distribution.
}
\label{realpofy}
\end{figure}

We test this assertion using the two-dimensional double-well potential
\[
V(x,y) = (x^2-1)^4 + x^2 y^2 + \mu y^2 + \nu y + \sigma x^2 y
\]
whose dynamics are predominantly chaotic near the saddle energy for 
appropriate parameter values and which 
we will later use to test tunnelling rate statistics. In practice we form 
statistical ensembles by fixing the energy and quantising $\hbar$ and in 
this way classical-dynamical parameters are held fixed throughout the 
spectrum. Our system does not allow the accumulation of sufficient 
data to test the distributions meaningfully for a fixed value
of $\theta=S(E)/\hbar$ and we must instead pass straight to a distribution 
in which all states are combined, of the kind illustrated in
Fig.~\ref{pxks}. We show in Fig.~\ref{realpofy}$(a)$ the accumulated
distribution $p(x_0)$ of overlaps with a harmonic oscillator ground state
(with $k=0$) for the parameter values $(\mu,\nu,\sigma) = (0.1,0,0)$. 
The potential in this case has a symmetry of reflection in $y$ and 
therefore has a real periodic orbit connecting to a tunnelling orbit in 
the $x$-axis. The harmonic oscillator used is the one which 
generates the monodromy matrix $W$ of the tunnelling orbit and from 
which the tunnelling operator is constructed. The distribution derived 
from the Gaussian hypothesis describes the quantum data  well
and indicates that it is a reasonable basis from which to tackle 
tunnelling-rate statistics.

To describe tunnelling-rate statistics, it is convenient to begin
once again with an ensemble taken from a fixed part of the spectral envelope
(with $\theta=S(E)/\hbar$ fixed mod~$2\pi$). Repeating the derivation 
of (\ref{localmom}), with $B_{lk}$ relaced by the matrix $\Tp$ representing 
the scaled tunnelling operator, we obtain
\[
p(y;\theta) = \frac{1}{2\pi} \int_{-\infty}^{\infty} e^{-iqy}
	       \tilde{p}(q,\theta)\d q,
\]
where 
\begin{equation}\label{tunresult}
\tilde{p}(q,\theta) =  \left[\det\ph{'\!\!}\left(I-\frac{2iq}{\beta} 
\Tp C(\theta)\right)
					\right]^{-\beta/2}.
\end{equation}
Our numerical calculations are made for GOE systems, in which case
$\beta=1$. 
(Note that while the mechanics of the calculation are similar to those 
leading to (\ref{localmom}), the quantities under investigation are
very different in each case.)
This distribution describes an ensemble of tunnelling 
rates taken with a fixed value of $\theta$.
As was the case for overlap statistics, it is not possible 
with our present model to produce enough data to test the distribution 
for narrow ranges of $\theta$ and instead we make a comparison in 
which states from all parts of the spectral envelope are combined. 
Theoretically, we produce a combined distribution describing  
such collective statistics by averaging $p(y,\theta)$ 
over $0\leq\theta<2\pi$.

Comparison of the resulting distribution with quantum data for the 
potential $V(x,y)$, with the same parameter values as in 
Fig.~\ref{realpofy}$(a)$, is shown in Fig.~\ref{realpofy}$(b)$ as a 
solid curve. Agreement with the scarred distribution is good and 
there is marked deviation from the prediction of standard RMT 
(short-dashed curve). Also shown for comparison is the  Porter-Thomas 
distribution \cite{BKH} (long-dashed curve).

Systems with symmetry such as that used for  Figs.~\ref{realpofy}$(a)$
and \ref{realpofy}$(b)$ are expected generically to show scarred 
statistics of the kind described in this paper. Nonsymmetric systems 
may also have scarred statistics, however, if parameters are tuned so that the 
real extension of the tunnelling orbit is periodic. We show tunnelling
rate statistics in  Figs.~\ref{realpofy}$(c)$ and \ref{realpofy}$(d)$
for one such potential, corresponding to the parameter values 
$(\mu,\nu,\sigma) = (0.25,0.40,0.254)$. Quantum data for this potential
were used in \cite{ourstats} to illustrate that dynamics could affect 
tunnelling statistics and the present theory gives a quantitative 
explanation of the qualitative observations made there. In  
Fig.~\ref{realpofy}$(c)$ we compare the scarred $p(y)$ 
(solid curve) with the standard RMT prediction
(short-dashed curve) and the Porter-Thomas distribution 
(long-dashed curve).

Note that while the scarred distribution appears
not to differ strongly from the Porter-Thomas distribution in 
Figs.~\ref{realpofy}$(b)$ and \ref{realpofy}$(c)$ , the difference 
is significant when plotted on a logarithmic scale. This is done for 
the nonsymmetric potential in Fig.~\ref{realpofy}$(d)$ (and omitted for 
the symmetric potential in the interests of space). We also note that
in the case of the symmetric potential the first hundred or so states
(omitted from Figs.~\ref{realpofy}$(a)$ and \ref{realpofy}$(b)$) give a 
distribution that seems better described by Porter-Thomas, though the 
scarred distribution unambiguously describes better the more highly
excited states corresponding. This semiclassical limit seems to be
approached more rapidly for the nonsymetric potential. 

We note that while the detailed discussion has been presented here for 
two-dimensional systems, the general theory will also work in 
higher dimensions. In particular, for example, (\ref{tunresult})
holds except that the matrices $\Tp$ and $C(\theta)$ have 
additional structure. Generalisation of $\Tp$ to higher dimensions 
is straighforward (see \cite{ourAP}) and, for $C(\theta)$, the 
generation of the autocorrelation function as described in the 
appendices may also be extended, though there are now more than 
the three classical parameters $(\Tr W,\Tr M_0,Q)$,  
controlling quantum recurrence.

The success of scarring theory in describing the tunnelling-rate
distribution therefore confirms the hypothesis of \cite{ourstats} 
that dynamical details of such low-dimensional systems can dominate 
the statistics of tunnelling and be responsible for strong deviation 
from RMT. The deviations we have seen are expected to be common in systems
with discrete symmetry, such as the double well potential we have 
treated in Figs.~\ref{realpofy}$(a)$ and \ref{realpofy}$(b)$.  Deviation 
will also be seen in more generic systems, however, as parameters are 
tuned to make the real extension of the tunnelling 
orbit periodic \cite{ourstats}. Such sensitive dependence of 
experimentally-accessible data on system parameters and dynamics
might be useful, for example, as a means of probing  the 
internal dynamics of relatively complex systems.

\section{Conclusion}\label{conclusion}
We have shown that wavefunction statistics in the neighbourhood
of a fixed point or short periodic orbit are well described by a 
normal distribution which allows for correlations between wavefunction 
components along different basis elements. We have also shown that this 
wavefunction distribution can be used to describe scarring anomalies 
previously detected in tunnelling-rate statistics.

Scarring effects lead to strong deviations from the standard
distributions of RMT in the neighbourhood of short periodic orbits.
To characterise these deviations fully, we examined the components 
of the eigenstates of a chaotic map in the eigenbasis of a harmonic 
oscillator centred on a periodic orbit. The low-lying 
states of such an eigenbasis are localised around the corresponding
periodic orbit and overlaps with a chaotic wavefunction are
strongly affected by scarring. In appropriate subintervals 
of the chaotic spectrum, we have proposed that the joint probability distribution 
describing the corresponding components of chaotic eigenstates remain
normal as in standard RMT but has a nontrivial covariance matrix (whose 
elements are completely determined by linearised dynamics around the 
periodic orbit). This hypothesis suffices to describe accurately all the 
measures of deviation from RMT that we have examined. The collective 
statistics of chaotic wavefunctions taken from a complete chaotic spectrum
are described by superimposing these distributions.
Deviations from RMT remain strong (and are effectively reproduced 
by the theory) in these combined ensembles, which 
are more natural to work with from a practical point of view.

We believe that examining the wavefunction statistics in a complete 
basis such as this provides a promising technique for understanding scarring 
in general terms. The description is essentially basis-independent and 
also completely determines the nature of wavefunction statistics
near a periodic orbit.

The assumption of normal deviation from RMT also allows the effect of 
scarring to be incorporated very naturally into the calculation of 
tunnelling rate statistics. The theory was tested on a two-dimensional 
double-well potential with chaotic dynamics and found to describe the 
statistics of its eigenfcuntions and energy-level splittings well. Scarring 
can account for quite strong deviation from standard RMT in this system. In 
this way we might use tunnelling and the statistics of tunnelling to probe 
sensitively the internal dynamics of unstable systems. 
\\

\vspace{1 cm}

\noindent {\bf Acknowledgements}\\
\\
\noindent SC Creagh and SY Lee are supported by the EPSRC under the 
Fast Stream scheme. SY Lee also acknowledges support from KOSEF.

\appendix

\section{generating the correlation function}\label{appgetA}
\noindent
We calculate the correlation function $A_{lk}(t)$ defined in 
section~\ref{overlapsec} using a generating function 
\begin{eqnarray}\label{defgen}
G(w,z,t) = \Tr\, e^{w\a} \T(z) \Uh_\lin^t,
\end{eqnarray}
where $\a$ is an annihilation operator for the harmonic oscillator
$\hh$ and 
\[
\T(z) = e^{(\ln z)\hh/\hbar}, 
\]
reduces to the tunnelling operator when $z=e^{-\alpha_0}$. Each of the 
three operators in this sequence is the quantisation of either
a linear symplectic transformation or a phase-space translation, 
possibly complex. The unitary operator $\Uh_\lin^t$ quantises
a real symplectic matrix $M_0^t$ which linearises the real unstable 
motion near $\zeta_0$. The operator  $\T(z)$, which is nonunitary, 
quantises the complex symplectic matrix $W(z)$ defined in (\ref{defW}).
Finally, the operator $e^{w\a}$ can be interpreted
as a translation along the complex phase-space displacement
\begin{equation}\label{defZ}
\delta \zeta = -w\sqrt{\frac{\hbar}{2}}\spinor{1}{i}.
\end{equation}
The combined operator therefore quantises the affine transformation
\[
\zeta \mapsto W(z)M_0^t\zeta + \delta \zeta.
\]

Because the transformation is affine, the trace formula may be used 
to evaluate the right hand side (\ref{defgen}) exactly. It is shown in 
appendix~\ref{appgetrace}  that  this gives
\begin{equation}\label{trgen}
G(w,z,t) = \frac{1}{\sqrt{\Delta(z,t)}} 
\exp\left[(-1)^{\mu t}\frac{iw^2}{2}\frac{z}{\Delta(z,t)}\,\tan\psi(t)\right]
\end{equation}
where
\[
\Delta(z,t) = \Tr\, W(z)M_0^t - 2 = 
(-1)^{\mu t}\left(m(t)z + \frac{m^*(t)}{z}\right) - 2
\]
and $m(t)=e^{i\phi(t)}\sec\psi(t)$ as described in Sec.~\ref{overlapsec}.
Note that $G(w,z,t)$ is independent of $\hbar$. Note also that
since $\Delta(z,t)$ is complex, care must be taken in deciding the
branch of the square root in (\ref{trgen}) --- the branch is 
chosen using the Maslov index (computed from real dynamics), so 
that $\Delta(z,t)$ reduces to the standard formula for real maps 
when $z\to 1$.

Expanding the right hand side of (\ref{trgen}) gives
\[
G(w,z,t) =\sum_{n=0}^\infty (-1)^{n\mu t}\frac{i^nw^{2n}}{2^n n!}
\sum_{k=0}^\infty z^{k+2n+1/2}\sin^n\psi \sqrt{\cos\psi} \;
C_k^{n+1/2}(\cos\psi)\;e^{i(n+k+1/2)(\phi-\mu t\pi)},
\]
where $C_k^{n+1/2}(\cos\psi)$ are Gegenbauer polynomials defined by 
the  generating function
\[
\frac{1}{(1-2xz+z^2)^{n+1/2}} = \sum_{k=0}^\infty C_k^{n+1/2}(x)z^k.
\]
Comparing this with the expansion
\begin{equation}\label{qmgen}
G(w,z,t) = \sum_{k=0}^\infty \sum_{m=0}^\infty \frac{w^m}{m!}\,
z^{k+m+1/2}\,\braopket{\kt}{\a^m \Uh_\lin^t}{\kt}
\end{equation}
obtained by evaluating the trace (\ref{defgen}) in an eigenbasis of $\hh$
leads us to the result
\[
\braopket{\kt}{\a^{2n}\Uh_\lin^t}{\kt} = \frac{(2n)!}{2^n n!}
\sin^n\psi \sqrt{\cos\psi} \; C_k^{n+1/2}(\cos\psi)\;
e^{i(k+1/2)(\phi-\mu t\pi)+in\phi+in\pi/2}
\]
from which we determine $A_{k+2n,k}^\lin(t)$ as given in (\ref{givecorr}). 
Note that we find
$A_{lk}^\lin(t)=0$ unless $l$ and $k$ are both even or both odd. 
This is because the linearised system enjoys an inversion symmetry even 
when the original system does not.

\section{tracing the affine transformation}\label{appgetrace}
\noindent
In this appendix we derive (\ref{trgen}) using the trace 
formula \cite{tabor}. We 
consider first the general affine transformation
\begin{equation}\label{realaffine}
\zeta\mapsto M\zeta+Z
\end{equation}
where $M$ and $Z=(a,b)$ are an arbitrary  symplectic matrix
and phase-space displacement respectively. For convenience
we may initially suppose that $M$ and $Z$ are real and
generalise to complex transformations later using analytic 
continuation.

The transformation in (\ref{realaffine}) has a single fixed point 
\[
\zeta = -(M-I)^{-1} Z.
\]
The trace formula then gives \cite{tabor}
\[
\Tr\, \hat{T}(Z) \Uh(M) = \frac{e^{iS_0/\hbar}}{\sqrt{\Tr\,M-2}},
\]
where $\hat{T}(Z)$ and $\Uh(M)$ are the operators which quantise 
$\zeta\mapsto\zeta+Z$ and $\zeta\mapsto M\zeta$ respectively and $S_0$ is 
the action of the fixed point.
By evaluating the generating function for $M$ at the fixed point
one can show that
\begin{eqnarray*}
S_0 &=& \ha(bq-ap) \\
&=& \ha\Omega(\zeta,Z) \\
&=& \ha\Omega(Z,(M-I)^{-1}Z),
\end{eqnarray*}
where $\Omega(u,v)=u^TJv$ is the symplectic form and $\zeta=(q,p)$.

We specialise to the trace in (\ref{defgen}) by substituting
the $\delta\zeta$ defined in (\ref{defZ}) for $Z$ and $W(z)M_0^t$
for $M$, giving
\[
G(w,z,t) =  \frac{e^{iw^2\Xi(z,t)/4}} {\sqrt{\Tr\,W(z)M_0^t-2}},
\]
where
\begin{equation}\label{defXi}
\Xi(z,t) = \Omega\left(\zeta_c,\;\frac{1}{W(z)M_0^t-I}\,\zeta_c\right) 
\quad\mbox{and}\quad
\zeta_c = \spinor{1}{i}.
\end{equation}
Notice that, because $\delta\zeta$ scales as $\hbar^{1/2}$, there is 
no $\hbar$-dependence in the final result.

We are free to work in a coordinate system in which $\Khat=I$ 
and then
\[
W(z) = e^{i \ln z J} 
=   z(I+iJ) +     \frac{1}{z}(I-iJ).
\]
Consider first the case
\begin{equation}\label{specM}
M_0^t = (-1)^{\mu t}\mat{\cosh\rho t}{e^\sigma\sinh\rho t}
{e^{-\sigma}\sinh\rho t}{\cosh\rho t}.
\end{equation}
where $\sigma$ and $\rho$ are constants. Substitution in (\ref{defXi})
gives then 
\[
\Xi(z,t) = 2 (-1)^{\mu t}\cosh\sigma  \,\sinh\rho t \;\frac{z}{\Delta(z,t)},
\]
where $\Delta(z,t)$ was defined in appendix A. Explicit calculation of 
$\Delta(z,t)$ for this $M_0$ gives
\[
Q = -\sinh\sigma.
\]
Using the identity
\[
\sec^2\psi(t) = \cosh^2\rho t + \sinh^2\sigma \sinh^2\rho t
= 1+\cosh^2\sigma \sinh^2\rho t
\]
we then get $\tan\psi(t)=\cosh\sigma \,\sinh\rho t$ and hence (\ref{trgen}).

Any hyperbolic matrix can be put in the form
(\ref{specM}) following an orthogonal transformation which
rotates coordinates by an angle $\theta_0$. The matrix $W(z)$
is unchanged by such a transformation. However, the annihilation 
operator is multiplied by a phase 
\[
\a\to e^{i\theta_0}\a
\]
and the exponent of the generating function by
\[
\Xi(z,t)\to e^{2i\theta_0}\Xi(z,t).
\]
In systems with time-reversal symmetry we can demand that
the eigenvectors $\ket{\kt}$ be real and this allows us to fix the 
possible rotation angles $\theta_0$ to multiples of $\pi/2$.
Among these, only $\theta_0=0$ and $\theta_0=\pi/2$ give distinct 
results for $\Xi(z,t)$.  The formula we have given applies to the 
case $\theta_0=0$. In the case $\theta_0=\pi/2$, the formula holds as 
given if we rephase the eigenvectors according to 
$\ket{\kt}\to(-1)^k\ket{\kt}$.

We can similarly use a convention for $\ket{\kt}$ such that
(\ref{givecorr}) holds in the GUE case. This amounts in particular to
adopting a convention in which the covariance matrix
$C_{lk}(E)=\langle x_l x_k^*\rangle$ is real even though individual
values of $x_l x_k^*$ are complex.  This reflects the fact that even
in systems without global time-reversal symmetry, the linearised
dynamics from which $A_{lk}^\lin(t)$ is calculated is time-reversal
invariant and we are free to choose the eigenstates $\ket{\kt}$ in a
way which reflects this. This is in direct analogy to the spatial
symmetry mentioned in the previous appendix which the linearised 
dynamics also possesses even in the absence of an equivalent symmetry 
for the global dynamics.

\end{document}